\documentclass{ws-ijbc}
\usepackage{ws-rotating}     % used only when sideways tables/figures are used
\usepackage{graphicx}
\usepackage{epstopdf}
\graphicspath{{./Figures/}}
\begin{document}

\catchline{}{}{}{}{} % Publisher's Area please ignore

\markboth{A. I. Ahamed and M. Lakshmanan}{Sliding Bifurcations in Memristive MLC Circuit and Chua Oscillator}

\title{SLIDING BIFURCATIONS IN THE MEMRISTIVE MURALI-LAKSHMANAN-CHUA CIRCUIT AND THE MEMRISTIVE DRIVEN CHUA OSCILLATOR}

\author{A. ISHAQ AHAMED}

\address{Department of Physics, Jamal Mohamed College,\\ Tiruchirappalli-620020,India\\
ishaq1970@gmail.com}

\author{M. LAKSHMANAN}
\address{Department of Nonlinear Dynamics,School of Physics,\\
Bharathidasan University,\\Tiruchirappalli-620024, India\\
lakshman@.gmail.com}

\maketitle

\begin{history}
\received{(to be inserted by publisher)}
\end{history}

\begin{abstract}
In this paper we report the occurrence of sliding bifurcations admitted by the memristive Murali-Lakshmanan-Chua circuit \cite{icha13} and the memristive driven Chua oscillator \citep{icha11}. Both of these circuits have a flux-controlled active memristor designed by the authors in 2011, as their non-linear element. The three segment piecewise-linear characteristic of this memristor bestows on the circuits two discontinuity boundaries, dividing their phase spaces into three sub-regions. For proper choice of parameters, these circuits take on a degree of smoothness equal to one at each of their two discontinuities, thereby causing them to behave as  \textit{Filippov} systems. Sliding bifurcations, which are characteristic of Filippov systems, arise when the periodic orbits in each of the sub-regions, interact with the discontinuity boundaries, giving rise to  many interesting dynamical phenomena. The numerical simulations are carried out after incorporating proper zero time discontinuity mapping (ZDM) corrections. These are found to agree well with the experimental observations which we report here appropriately.
\end{abstract}


%\begin{multicols}{2}
\section{Introduction}

Among the nonlinear systems, there are large classes of systems called  \textit{non-smooth systems} or \textit{piecewise-smooth systems}. These system are found to contain terms that are non-smooth functions of their arguments and fall outside the purview of the conventional theory of dynamical systems. Examples of such systems are electrical circuits which have switches, mechanical devices wherein components impact against each other and many control systems where continuous changes may trigger discrete actions. These systems have their phase-space divided into two or more sub-regions by the presence of what are called  \textit{discontinuity boundaries}. They are characterized by functions that are \emph{event driven}, that is they are normally smooth but lose their smoothness at the discontinuity due to instantaneous events such as the application of a switch \cite{dib08}. Extensive studies on piecewise-smooth systems were made by \citet{brog99,brog2k},  \citet{kunze2k},  \citet{mose03} and \citet{leine04}. Earlier studies of these piecewise-smooth systems can also be found in the Eastern European literature, particularly the the pioneering work of \citet{adnov65} on non-smooth equilibrium bifurcations, Feigin's work on \emph{C-Bifurcations} \cite{feigin94,dib_csf99}, \citet{peterka74} and \citet{bab78} works on impact oscillators and \citet{flip88} work on sliding motions. All of these studies reveal that these systems posses rich and complex dynamics. 

The piecewise-smooth systems, in addition to the familiar dynamical behaviours exhibited by smooth dynamical systems, are found to exhibit what are called as Discontinuity Induced Bifurcations (DIB's). Examples of these are the \textit{border-collision bifurcations} \citep{feigin70,nusse92}, \textit{boundary equilibrium bifurcations} \citep{dib02a,leine02,leine04}, \textit{grazing bifurcations} \citet{nord91,dib01}, \textit{sliding and sticking bifurcations} \citet{feigin94,dib_ijbc01,dib02,dib_ijbc03}, \textit{boundary intersection bifurcations} \cite{nusse92} etc. The piecewise-smooth systems are classified as \textit{piecewise-smooth maps}, \textit{piecewise-smooth continuous} or \textit{hybrid piecewise-smooth systems} based on their dependence on time. The piecewise-smooth systems having a continuous time dependence are further classified as \textit{Filippov systems} or \textit{Flows} based on the order of their discontinuity. Filippov systems have a discontinuity of order one \citep{flip64,flip88}, while the flows have an order of discontinuity two or greater than two. 

The memristive Murali-Lakshmanan-Chua (MMLC) circuit \cite{icha13} and memristive driven Chua oscillator \cite{icha11} have an active flux-controlled memristor designed by the present authors in 2011 as their nonlinear element. This flux-controlled memristor, by virtue of its nonlinear characteristic, endows the circuits with two discontinuity boundaries causing them to behave as piecewise-smooth systems. In their earlier studies on the memristive MLC cirucit \citep{icha13,icha17} the authors have shown that it is a continuous flow with a degree of smoothness equal to two admitting \textit{grazing bifurcations}, \textit{hyperchaos}, \textit{transient hyperchaos}, \textit{Hopf} and \textit{Neimark-Sacker bifurcations}. However for proper choice of parameters, both the memristive MLC circuit and the memristive driven Chua oscillator can have a uniform discontinuity with a degree of smoothness one, thereby causing them to become Filippov systems. In this paper we report the sliding bifurcations, characteristic to Filippov systems, admitted by these memristive circuits. The paper is organized as follows. In Sec. 2, we discuss the different types of \textit{sliding} bifurcations that occur in a n-dimensional system as well as the conditions for the occurrence of the same. In Sec. 3, we discuss the analog model of the memristor, its characteristic features and its experimental realization using off-the-shelf components. We then describe briefly the memristive MLC circuit, its circuit equations and their normalized forms using proper rescaling parameters. Further we consider the memristive MLC circuit as a nonsmooth system and the reformulation of the system's equations as a set of smooth odes. We then derive the conditions for the occurrence of the \textit{sliding} bifurcations, and the Zero Time Discontinuity Mapping (ZDM) corrections that are required for observing them numerically. In Sec. 4 the sliding bifurcations induced dynamics in the memristive MLC circuit is analyzed both numerically and through hardware experiments. Similar to the description of the memristive MLC circuit, in Sec. 5 we describe the memristive driven Chua oscillator, list its circuit equations and their normalized forms and reformulate the circuit as a nonsmooth system. In Sec. 6 we derive the conditions for occurrence of \textit{sliding} bifurcations and the ZDM corrections required for observing them in the memristive Chua's circuit. Further we report the \textit{crossing-sliding} and \textit{grazing-sliding} bifurcations occurring in it using numerical simulations as well as laboratory experiments. In Sec. 7 we present the results and conclusions.

\section{Sliding Bifurcations in a General $n$-Dimensional System}
Sliding bifurcations are Discontinuity Induced Bifurcations (DIB's)
arising due to the interactions between the limit cycles of a Filippov system  with the boundary of a sliding region. Four types of sliding bifurcations have been identified by Feigin \cite{feigin94} and were subsequently analyzed by di Bernado, Kowalczyk and others \cite{dib_ijbc01,kowal01,dib02,dib_ijbc03} for a general $n-$dimensional system. These four sliding bifurcations are \emph{crossing-sliding} bifurcations, \emph{grazing-sliding} bifurcations, \emph{switching-sliding} bifurcations and \emph{adding-sliding} bifurcations. In this section we describe briefly the various types of sliding bifurcations, the conditions for their occurrence and the Zero Time Discontinuity Mapping (ZDM) corrections that are essential for observing these numerically. 

Let us consider a piecewise-smooth continuous system having a single discontinuity boundary $\Sigma_{i,j}$. Let this discontinuity boundary divide the phase space into two sub-regions $S_1$ and $S_2$. Further let this discontinuity boundary be \emph{uniformly discontinuous} in some domain $\mathcal{D}$. This means the degree of smoothness of the system is the same for all points $x \in \Sigma_{i,j} \cap \mathcal{D}$.  Then this system can be written as the zero set of a smooth function $H$ such that 
\begin{equation}
\dot{x} = \left \{
					\begin{array}{ll}
						F_1(x,\mu) \qquad \qquad \textrm{if}\qquad H(x) > 0 \\
						F_2(x,\mu) \qquad \qquad \textrm{if}\qquad H(x) < 0.
					\end{array}
				\right.
		\label{eqn:flip}
\end{equation}
where $F_1(x) \neq F_2(x)$ at $H(x) = 0$. 

\subsection{Types of Sliding Bifurcations}
\begin{figure}[htbp]
	\centering
	\resizebox{0.8\textwidth}{!}
		{\includegraphics{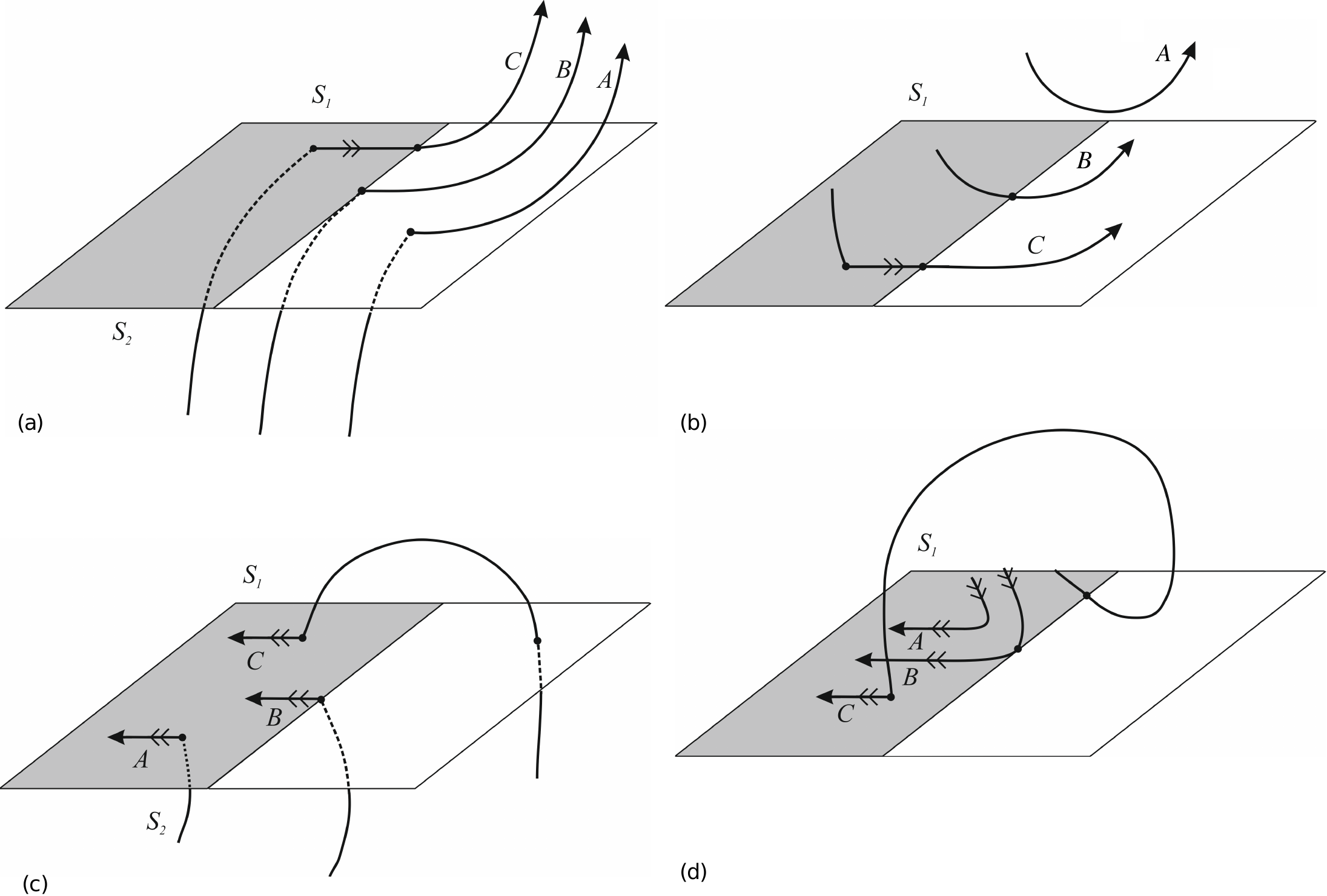}}
		\rule{35em}{0.5pt}
	\caption[Types of sliding bifurcations] {The four types of sliding bifurcations (a) crossing-sliding (b) grazing-sliding bifurcations (c) switching-sliding and (d) adding-sliding bifurcation.}
	\label{fig:slide_cases}
\end{figure}

To analyze the sliding bifurcations that may occur in this general $n$-dimensional system, the state equations, Eqs. (\ref{eqn:flip}), are formulated using {\emph{Utkin's Equivalent Control Method}} (\citet{utkin92}). In this method it is assumed that the system flows according to a sliding vector field, say $F_S$ which is the average of the two vector fields $F_1$ in region $S_1$ and $F_2$ in region $S_2$ plus a control $\beta(x)\in[-1,1]$ in the direction of the difference between the vector fields.
\begin{equation}
F_s = \dfrac{F_1+F_2}{2}+\frac{F_2-F_1}{2}\beta(x),
	\label{eqn:utkins_slide} 
\end{equation}
where the equivalent control is specified as
\begin{equation}
\beta(x) = -\dfrac{H_xF_1+H_xF_2}{H_xF_2-H_xF_1}.
	\label{eqn:utkins_beta}
\end{equation}
The sliding region is given by 
\begin{equation}
\widehat{ \Sigma} := \{ x \in \Sigma : -1 \leq \beta \leq 1 \},
	\label{eqn:utkins_sigma}
\end{equation}
and its boundaries are 
\begin{equation}
\partial \widehat{\Sigma}^{\pm} := \{ x \in \Sigma: \beta = \pm 1 \}.
	\label{eqn:utkins_sigma_bound}
\end{equation}
A schematic of all the sliding bifurcations are depicted in Fig. \ref{fig:slide_cases}. Generally we assume three trajectories, say $A,\,B$ and $C$ in the neighbourhood of the switching boundary $\widehat{\Sigma}$ and study their behaviours in the sliding region bounded by $\partial \widehat{\Sigma}^{\pm}$. Of these we consider trajectory $B$ as a \emph{fiducial} trajectory that crosses the sliding boundary at a precise boundary equilibrium point.

\begin{enumerate}
\item
{\textbf{Crossing-Sliding Bifurcations:}}\\
The incidence of \emph{crossing-sliding} bifurcation is shown in Fig. \ref{fig:slide_cases}(a). Here, while all the trajectories cross the switching manifold transversely, the fiducial trajectory $B$ forms the boundary between two topologically different trajectories, with the trajectory $C$ undergoing a sliding motion in a small segment of the sliding region and trajectory $A$ not exhibiting any sliding motion at all. The trajectory $B$ therefore heralds the onset of sliding motion. This bifurcation is so called because all the trajectories incident on the switching manifold $\widehat{\Sigma}$ with zero speed, but as we move from $A \rightarrow B \rightarrow C$, we \emph{cross} the sliding boundary $\partial \widehat{\Sigma}$.

\item
{\textbf{Grazing-Sliding Bifurcations:}}\\
Here the scenario is similar to the grazing bifurcations encountered in general piecewise-smooth continuous flows having a higher degree of smoothness, hence called as \emph{grazing-sliding} bifurcations. This is shown in Fig. \ref{fig:slide_cases}(b). Here a trajectory $A$ lying entirely in region $S_1$ when perturbed continuously gets converted into the trajectory $B$ which forms a point of \emph{grazing} with the switching manifold. Under further perturbation, it gets converted into the trajectory $C$ which slides in a small section of sliding region, crosses the boundary $\partial\widehat{\Sigma}$ and leaves $\Sigma$ transversely.

\item
{\textbf{Switching-Sliding Bifurcations:}}\\
Here as we perturb the trajectory in the sequence $ A \rightarrow B \rightarrow C $, we have an extra transversal crossing of $\Sigma$ or an extra \emph{switching} transition. This is shown in Fig. \ref{fig:slide_cases}(c). This is different from \emph{crossing-sliding} bifurcation in that the sliding boundary $\partial\widehat{\Sigma}$ is repelling within the sliding region, whereas it is attractive in the \emph{crossing-sliding} case.

\item
{\textbf{Adding-Sliding Bifurcations:}}\\
In this case, for a transition $A \rightarrow B \rightarrow C$, a locally uninterrupted sliding motion is transformed into two separate pieces of sliding, separated by a region of free non-sliding evolution. This causes an \emph{addition} of one sliding segment to the general motion of the system under perturbation, and hence the name. This type is shown in Fig. \ref{fig:slide_cases}(d)

\end{enumerate}
\subsection{Conditions for the Existence of Sliding Bifurcations}
Following \citet{dib08}, the conditions for the onset of various types of sliding bifurcations are given. Firstly the defining condition for all sliding motions is 
\begin{equation}
H_xF_1(x^*)=0.
	\label{eqn:slide_cond1}
\end{equation}
This condition is to ensure that all the trajectories impinge on the switching manifold $\hat{\Sigma}$ transversely. In addition to this there is a non-degeneracy assumption valid for all sliding bifurcations 
\begin{equation}
H_xF_d(x^*) >0,
	\label{eqn:slide_cond2}
\end{equation}
where $F_d = (F_2-F_1)$ is the difference between the two vector fields.  For \emph{crossing-sliding} and \emph{grazing-sliding} bifurcations we have an extra non-degeneracy condition, namely
\begin{equation}
H_xF_{1,x}F_1(x^*) >0.
	\label{eqn:slide_cond3}
\end{equation}
This extra non-degeneracy condition ascertains whether the boundary $\partial \widehat{\Sigma}$ is attracting or repelling to the sliding flow.
For the \emph{switching-sliding} bifurcation, this condition becomes
\begin{equation}
H_xF_{1,x}F_1(x^*) <0,
	\label{eqn:slide_cond4}
\end{equation}
while for \emph{adding-sliding} bifurcation it is 
\begin{equation}
H_xF_{1,x}F_1(x^*) =0.
	\label{eqn:slide_cond5}
\end{equation}
This condition is a defining condition for the existence of a point of tangency of the adding-sliding flow with $\partial \widehat{\Sigma}^-$ at the bifurcation point. Further for the \emph{adding-sliding} bifurcation an extra inequality condition has to be satisfied, namely
\begin{equation}
H_xF_{1,x}^2F_1(x^*)+H_xF_{1,xx}F_1^2(x^*) < 0.
	\label{eqn:slide_cond6}
\end{equation}

\subsection{ZDM Corrections for Sliding Bifurcations}

The concept of a discontinuity map was introduced by \citet{nord91}. This is a synthesized Poincar\'{e} map that is defined \textit{locally} near the point at which a trajectory interacts with the discontinuity boundary. When composed with a global Poincar\'{e} map, say around a limit cycle, ignoring the presence of the discontinuity boundary, then one can derive a non-smooth map whose orbits describe completely the dynamics of the system. Let $\Sigma_{12}$ be a discontinuity surface that separates the phase space of a dynamical system into two regions $S_1$ and $S_2$. Let the trajectory in the sub-region $S_1$ be described by the flow vector $\Phi_1(t)$. When this trajectory intersects the discontinuity boundary $\Sigma_{12}$ and crosses over to the sub-region $S_2$ it will be described by the flow vector $\Phi_2(t)$. For periodic motions, the system trajectory crosses the discontinuity boundary $\Sigma_{12}$ multiple number of times. If the total time elapsed between successive intersections is assumed to be zero, then the mapping of the periodic orbit local to the intersecting point that translates flow vectors $\Phi_1(t)$  to $\Phi_2(t)$ is the Zero Time Discontinuity Mapping.

The Zero Time Discontinuity Map (ZDM) corrections for all the four types of sliding bifurcations for a general $n-$dimensional Filippov flow described by Eqs. (\ref{eqn:flip}), refer \cite{dib08}, evaluated at the boundary equilibrium point $x = x^*$ are listed here.
\begin{enumerate}
\item
{\bfseries{Crossing-sliding:}}\\
For trajectories starting in the region $S_2$,$(H(x)<0)$ the ZDM correction for {\emph{crossing-sliding}} bifurcation is 
\begin{equation}
x \mapsto \left\{
				\begin{array}{ll}
					x \qquad\qquad\text{if}\,\,(H_xF_1)_x(x-x^*) \leq 0,\\ 
					x+\delta_1 \qquad \text{if}\,\,(H_xF_1)_x(x-x^*) > 0, \\
				\end{array}
			\right.
		\label{eqn:slide_zdm1}
\end{equation}
where 
\begin{equation}
\delta_1=\dfrac{1}{2}\dfrac{[(H_xF_1)_x(x-x^*)]^2}{(H_xF_2)[(H_xF_1)_xF_1]}F_d. \nonumber
\end{equation}
\item
{\bfseries{Grazing-sliding:}}\\
For trajectories starting in the region $S_1$,$(H(x)>0)$ the ZDM correction for {\emph{grazing-sliding}} bifurcation  is 
\begin{equation}
x \mapsto \left\{
				\begin{array}{ll}
					x \qquad\qquad \text{if}\,\,H_x(x-x^*) \geq 0, \\ 
					x+\delta_2 \qquad \text{if}\,\,H_x(x-x^*) < 0, \\
				\end{array}
			\right.
		\label{eqn:slide_zdm2}
\end{equation}
where 
\begin{equation}
\delta_2=-\dfrac{H_x(x-x^*)}{H_xF_2}F_d. \nonumber
\end{equation}
\item
{\bfseries{Switching-sliding:}}\\
For trajectories starting in the region $S_2$,$(H(x)>0)$ the ZDM correction for {\emph{switching-sliding}} bifurcation is 
\begin{equation}
x \mapsto \left\{
				\begin{array}{ll}
					x \qquad\qquad \text{if}\,\, (H_xF_1)_x(x-x^*) \leq 0, \\ 
					x+\delta_3 \qquad \text{if}\,\,(H_xF_1)_x(x-x^*) > 0, \\
				\end{array}
			\right.
		\label{eqn:slide_zdm3}
\end{equation}
where 
\begin{equation}
\delta_3=\dfrac{2}{3}\dfrac{[(H_xF_1)_x(x-x^*)]^3}{(H_xF_2)^2[(H_xF_1)_xF_1]^2}Q, \nonumber
\end{equation}
and
\begin{equation}
Q = [(H_xF_2)(F_{1,x}F_d-F_{d,x}F_1)-(H_x(F_{1,x}F_d-F_{d,x}F_1))F_d]. \nonumber
\end{equation}
\item
{\bfseries{Adding-sliding:}}\\
For trajectories starting in the region $S_1$,$(H(x)<0)$ the ZDM correction for {\emph{adding-sliding}} bifurcation is 
\begin{equation}
x \mapsto \left\{
				\begin{array}{ll}
					x \qquad\qquad \text{if}\,\, (H_xF_1)_x(x-x^*) \geq 0, \\ 
					x+\delta_4 \qquad \text{if}\,\,(H_xF_1)_x(x-x^*) < 0, \\
				\end{array}
			\right.
		\label{eqn:slide_zdm4}
\end{equation}
where 
\begin{equation}
\delta_4=-\dfrac{9}{2}\dfrac{[(H_xF_1)_x(x-x^*)]^2}{(H_xF_2)^2\{[(H_xF_1)_xF_1]_xF_1\}}Q, \nonumber
\end{equation}
and $Q$ is as defined earlier.
\end{enumerate}

\section{Memristive Murali-Lakshmanan-Chua Circuit} 

A memristive MLC circuit \cite{icha13} was designed by the authors by removing the Chua's diode and replacing it with an active flux-controlled memristor in the original Murali-Lakshmanan-Chua circuit. A memristor can be defined as any two-terminal device which exhibits a pinched hysteresis loop in the $(v-i)$ plane when driven by a bipolar periodic voltage or current excitation waveform, for any initial conditions, refer Chapter 2 by Leon O Chua in \cite{ron14}. Following the work on memrestive oscillators by \citet{itoh08}, a large number of researchers have proposed different models for the memristor.
\subsection{Analog Memristor Model}
\begin{figure}[htbp]
	\centering
	\resizebox{\textwidth}{!}
		{\includegraphics{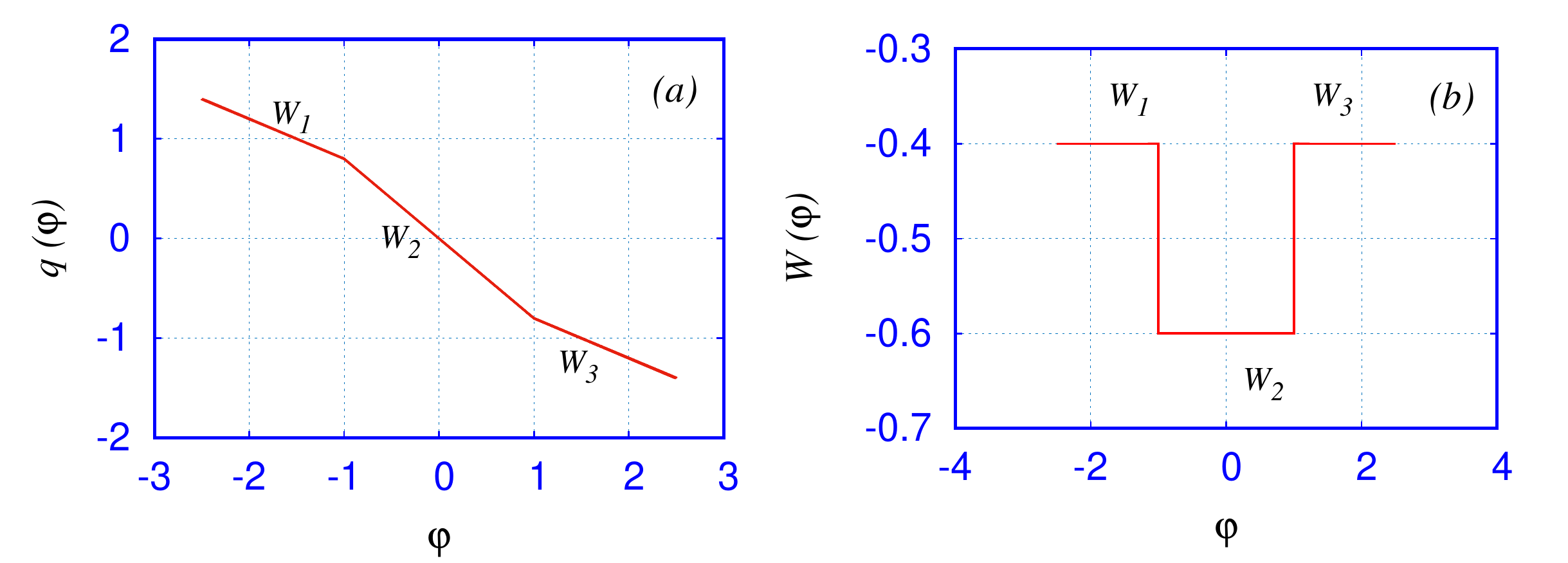}}
	\caption[Characteristic of the Active Flux Controlled Memristor] {Characteristic of the active flux controlled memristor :(a) in the $(\phi-q(\phi))$ plane and (b) variations of the memductance $W(\phi)$  as a function of the flux $\phi$.}
	\label{fig:mem_cha2}
\end{figure}

\begin{figure}[]
	\centering
	\resizebox{1.0\textwidth}{!}
		{\includegraphics{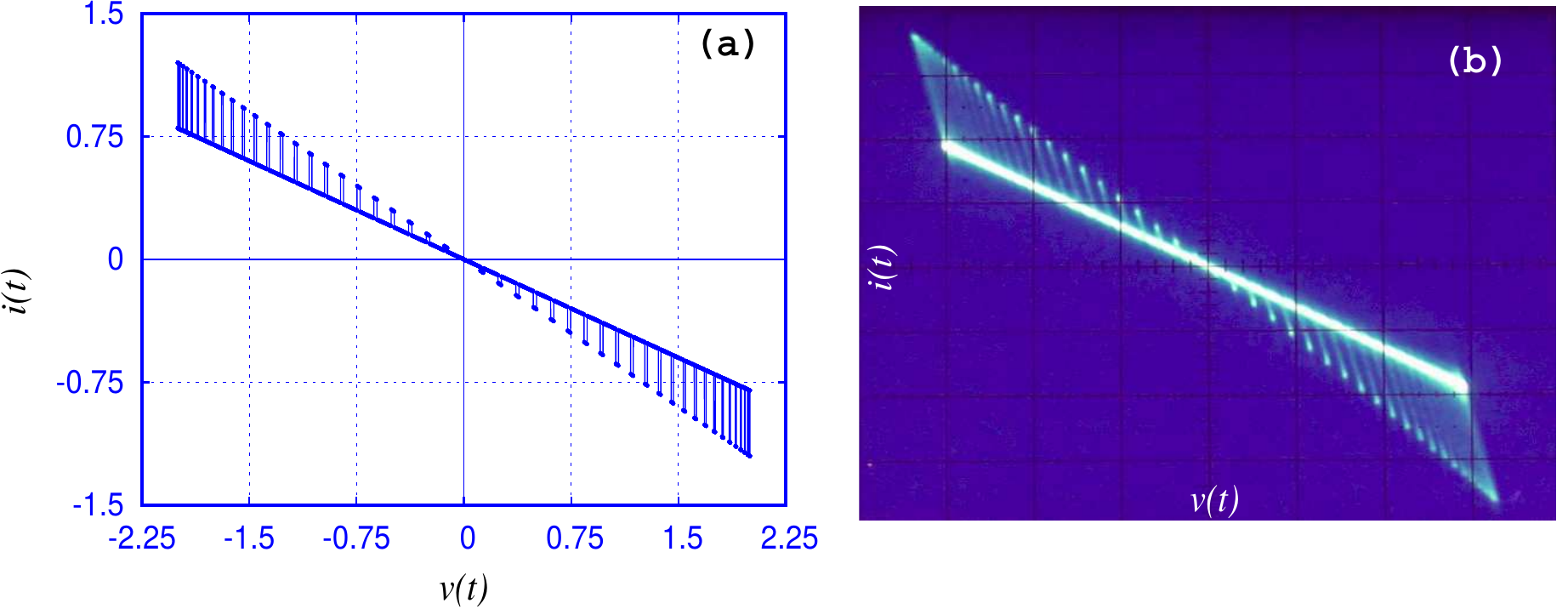}}
	\caption[Memristor Characteristic] { Memristor Characteristic (a) in the $(v(t)-i(t))$ plane obtained using numerical simulation and (b) the corresponding experimentally obtained characteristic.}
	\label{fig:Chac}
\end{figure}

The memristor model used in this circuit was introduced by the authors in 2011, \cite{icha11}. It has a three segment piecewise linear characteristic defined in the $(\phi-q)$ plane as shown in Fig. \ref{fig:mem_cha2}(a).
Mathematically the piecewise linear relationship is represented as
\begin{equation}
q(\phi)  =  W_2\,\phi + 0.5(W_1-W_2)[|\phi+1|-|\phi-1|],
	\label{eqn:flux_cont1}
\end{equation}
Here the charge $q(\phi)$ is given as a function of the flux $\phi$ across the memristor. As the charge $q(\phi)$ varies, the memductance $W(\phi) = \dfrac{\partial q(\phi)}{\partial \phi}$ takes on a value $W_2$ for $|\phi| \leq 1$ and a value $W_1 \equiv W_3 $ for $|\phi|>1$. Hence this is called a \emph{flux controlled memristor}. The variation in the memductance $W(\phi)$ as function of the variation of the flux $\phi$ across the memristor is shown in Fig. \ref{fig:mem_cha2}(b). Further the variations in the memductance values causes the resultant memductance $W(\phi)$ of the memristor to vary as a function of time $t$, with a low OFF state value of $W_2$ and high ON state value of $W_{1,3}$. This causes the current through the memristor to vary as a function of the voltage applied across it as 
\begin{equation}
i(t)  =  W_i(\phi)v(t) \qquad \qquad \mathrm{i = 1,2,3},
	\label{eqn:v-i}
\end{equation}
The $(v-i)$ characteristic obtained numerically by plotting the current $i(t)$ as a function of the memristor voltage $v(t)$ and is shown in Fig. \ref{fig:Chac}(a). This characteristic shows a continuous switching between a set of two straight lines with slopes equal to the two memductance values $W_2$ and $W_{1,3}$ as the memductance $W(\phi)$ varies from a low  OFF state to a high ON state and vice versa as time progresses. The corresponding experimentally observed characteristic in the $(v-i)$ plane is shown in Fig. \ref{fig:Chac}(b). It is to be noted that there is a good agreement between the numerical and experimental $(v-i)$  characteristic.\\

\begin{figure}[htbp]
	\centering
	\resizebox{\textwidth}{!}	
		{\includegraphics{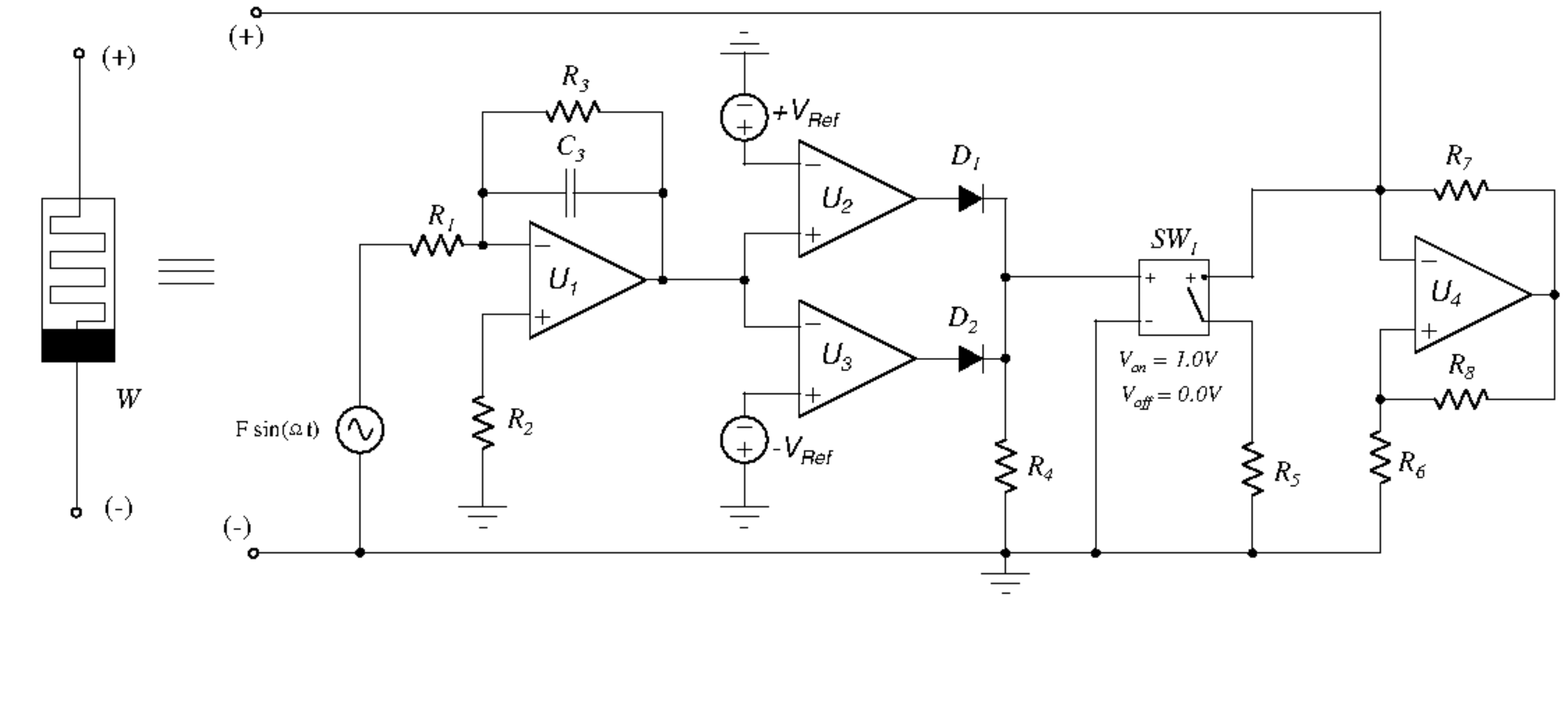}}
	\caption[Analog Circuit Model of the Memristor] {Analog Circuit Model of the Memristor}
	\label{fig:mem_pm}
\end{figure}

The circuit realization of this memristor is given in Fig. \ref{fig:mem_pm}.
The analog circuit of the memrsitor shown in Fig. \ref{fig:mem_pm} was implemented in the laboratory using off the shelf components. Here the integrator, the window comparator as well as the negative impedance converter were implemented using TL081 operational amplifiers. The electronic switch $AD7510DJ$ was used for realizing the switching action. The reason for using TL081 is that it operates at high frequency ranges, has high slew rate and does not show hysteretic behaviour. For the integrator part of the memristor model shown in Fig. \ref{fig:mem_pm}, the parameters were chosen as $ R_1 = 10 K\Omega$, $R_2 = 10 K\Omega$, $R_3 = 200 K\Omega$ and $C_3 = 2.2 nF$. For the window comparator part the output resistance was selected as $R_4 = 10 K\Omega$, while the reference voltages were fixed as $\pm 1V$. Further we had selected the linear resistances $R_5 = 3.3K\Omega$ and $R_6 = 1800 \Omega$, $R_7 = 2 K\Omega$ and $R_8 = 2 K\Omega$ for the negative conductance. The experimental observations were made by employing a Hewlett-Packard Arbitrary Function Generator (33120A) of frequency $15$ MHz, an Agilent Mixed Storage Oscilloscope (MSO6014A) of frequency $100$ MHz and sampling rate of $2$ Giga Samples / seconds.

\subsection{Experimental Realization of Memristive Murali-Lakshmanan-Chua Circuit}
\begin{figure}[htbp]
	\centering
	\resizebox{0.65\textwidth}{!}
		{\includegraphics{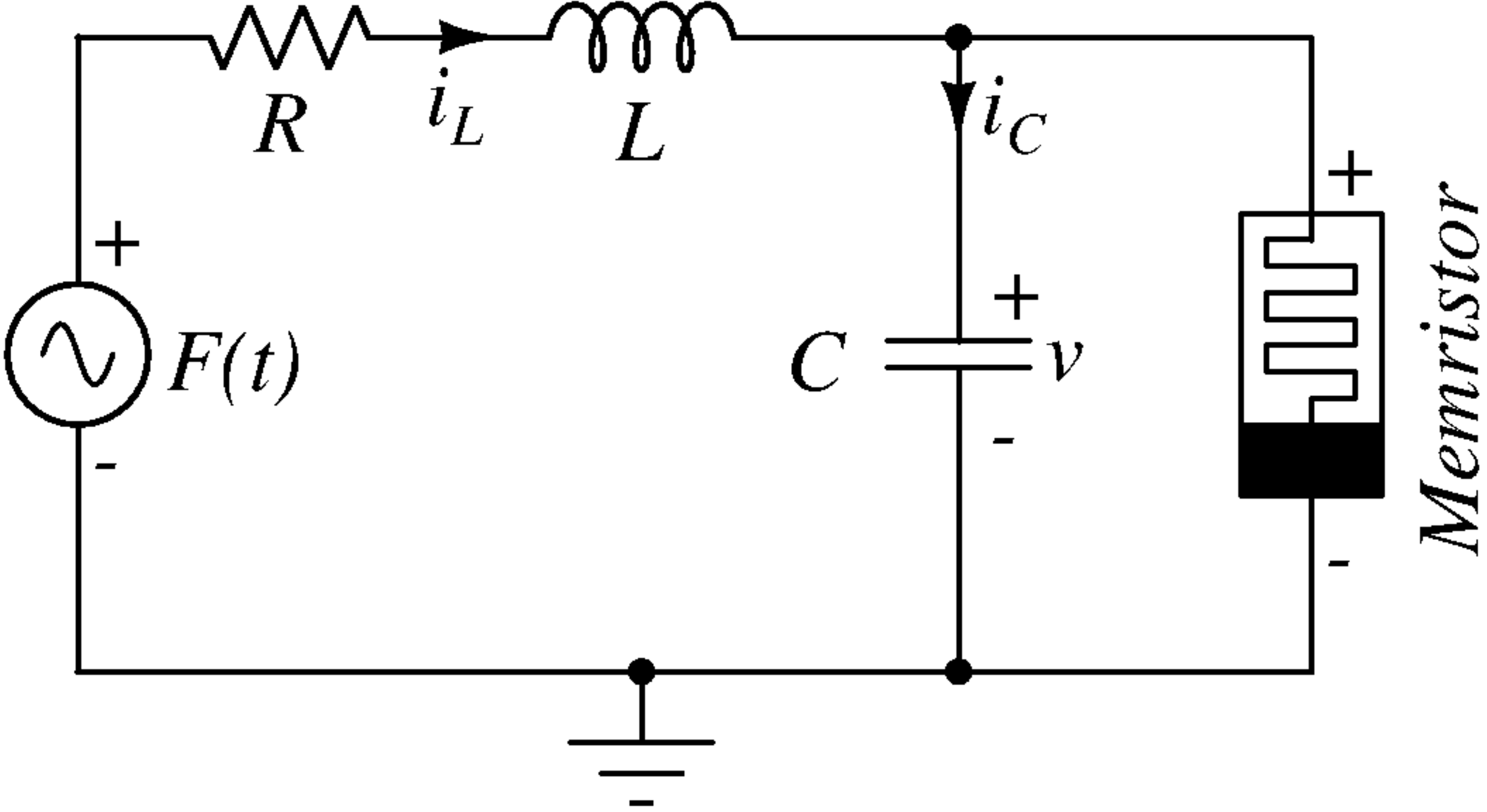}}
	\caption[The experimental realization of the memristive MLC Circuit] {The experimental realization of the memristive MLC circuit, with the active flux-controlled memristor part shown by the dashed outline. The parameter values of the circuit  are fixed as $ L = 21 mH$, $R = 900 \Omega$, $C_1 = 10.5nF$. The frequency of the external sinusoidal forcing is fixed as $\nu_{ext} = 8.288 kHz$ and the amplitude is fixed as $F = 770 mV_{pp} $ ( peak-to-peak voltage).}
	\label{fig:mmlc_cir}
\end{figure}

The experimental realization of the memristive MLC circuit is shown in Fig. \ref{fig:mmlc_cir}. Using Kirchhoff's current and voltage laws, the circuit equations can be written as a set of autonomous ordinary differential equations (ODEs), by taking the flux $\phi(t)$, voltage $v(t)$, current $i(t)$ and the time $p$ as variables as
\begin{eqnarray}
\frac{d\phi}{dt}  & = & v, \nonumber \\
C\frac{dv}{dt}  & = & i - W(\phi)v,  \nonumber \\ 
L \frac{di}{dt}  & = &  -v -Ri +F\sin( \Omega p),\nonumber \\
\frac{dp}{dt}  & = & 1.
	\label{eqn:mlc_cir}
\end{eqnarray}
Here $W(\phi)$ is the memductance of the memristor and is as defined in \cite{itoh08},
\begin{equation}
W(\phi) = \frac{dq(\phi)}{d\phi} = \left\{
					\begin{array}{ll}
					W_1, ~~~  \phi  < 1  \\
					W_2, ~~~ | \phi  | \leq 1\\
					W_3, ~~~  \phi >1,	
					\end{array}
				\right.
	\label{eqn:W}
\end{equation}
where $W_1$ and $W_3$ are the slopes of the outer segments, such that $W_1 = W_3$ and $W_2$ is the slope of the inner segment of the characteristic curve of the memristor respectively, refer Fig. \ref{fig:mem_cha2}(a). 
These circuit equations, Eqs. (\ref{eqn:mlc_cir}), can be  normalized as
\begin{eqnarray}
\dot{x}_1  & = & x_2, \nonumber \\
\dot{x}_2  & = & x_3-W(x_1)x_2, \nonumber \\ 
\dot{x}_3  & = & -\beta(x_2+x_3) + f \sin(\omega x_4),\nonumber \\
\dot{x}_4  & = &  1.
\label{eqn:mlc_nor}
\end{eqnarray}
Here dot stands for differentiation with respect to the normalized time $\tau$ (see below) and $W(x_1)$ is the normalized value of the memductance of the memristor, given as
\begin{equation}
W(x_1) = \frac{dq(x_1)}{dx_1} = \left\{
		\begin{array}{ll}
		a_1, ~~~  x_1 < 1 \\
		a_2, ~~~ | x_1 | \leq  1 \\
		a_3, ~~~ x_1 > 1,
		\end{array}
	\right.
	\label{eqn:W_nor}
\end{equation}
where $ a_1 = W_1/G $ and $a_3 = W_3/G$  are the normalized values of $W_1 $ and $W_3 $ with $a_1 = a_3$, while $a_2 = W_2/G$ is the normalized value of $W_2$ mentioned earlier and are negative. The rescaling parameters used for the normalization of the circuit equations are
\begin{eqnarray}
x_1 = \frac{G\phi}{C},x_2 = v, x_3 = \frac{i}{G}, x_4 = \frac{Gp}{C}, G = \frac{1}{R},\beta = \frac{C}{LG^2}, \\ \nonumber
\omega = \frac{\Omega C}{G} = \frac{2\pi \nu C}{G},\tau = \frac{Gt}{C},f = F\beta.
	\label{eqn:rescale} 
\end{eqnarray}

In our earlier work on this memristive MLC circuit, see \citet{icha13}, we reported that the addition of the memristor as the nonlinear element converts the system into a piecewise-smooth continuous flow having two discontinuous boundaries, admitting \emph{Grazing bifurcations}, a type of discontinuity induced bifurcation (DIB). These grazing bifurcations were identified as the cause for the occurrence of hyperchaos, hyperchaotic beats and transient hyperchaos in this memristive MLC system. Further we reported the occurrence of discontinuity induced \textit{Hopf} and \textit{Neimark-Sacker} bifurcations in the same circuit \cite{icha17}. In this paper we report the occurrence of sliding bifurcations in this circuit. These diverse dynamical behaviours prove that this simple low dimensional circuit is a robust and versatile dynamical system.

\subsection{Memristive MLC Circuit as a Non-smooth System}

The memristive MLC circuit is a piecewise-smooth continuous system by virtue of the discontinuous nature of its nonlinearity, namely the memristor. Referring to the memductance characteristic, we find that the memristor switches states at $x_1 = +1$ and at $x_1 = -1$ either from a more conductive ON state to a less conductive OFF state or vice versa. These switching states of the memristor give rise to two discontinuity boundaries or switching manifolds, $\Sigma_{1,2}$ and $\Sigma_{2,3}$ which are symmetric about the origin and are defined by the zero sets of the smooth functions $H_i(\mathbf{x},\mu) = C^T\mathbf{x}$, where $C^T = [1,0,0,0]$ and $\mathbf{x}= [x_1,x_2,x_3,x_4]$, for $i=1,2$. Hence $H_1(\mathbf{x}, \mu) = (x_1-x_1^\ast)$, $x_1^\ast = -1$ and $H_2(\mathbf{x}, \mu) = (x_1-x_1^\ast)$, $x_1^\ast = +1$, respectively. Consequently the phase space $\mathcal{D}$ can be divided into three subspaces $S_1$, $S_2$ and $S_3$ due to the presence of the two switching manifolds. The memristive MLC circuit can now be rewritten as a set of smooth ODEs 

\begin{equation}
\dot{\mathbf{x}}(t) = 
	\left\lbrace	\begin{array}{lcccl}
		F_2(\mathbf{x},\mu),& H_1(\mathbf{x}, \mu) \geq  0& \& & H_2(\mathbf{x}, \mu) \leq 0,& \mathbf{x} \in S_2,  \\
				   &					&    &                 &           \\
		F_{1,3}(\mathbf{x},\mu), &  H_1(\mathbf{x}, \mu)< 0 &\& &H_2(\mathbf{x},\mu)> 0, &\mathbf{x} \in S_{1,3},
		\end{array}
	\right.
	\label{eqn:smooth_odes}
\end{equation}
where $\mu$ denotes the parameter dependence of the vector fields and the scalar functions. The vector fields $F_i$'s are

\begin{equation}
 F_i(x,\mu) =  \left (	\begin{array}{cccc}
				0	&x_2		&0			&0	\\
				0	&-a_i x_2	&x_3		&0	\\
				0	&-\beta x_2 &-\beta x_3 &0	\\	
				0	&0			&0			&1 
				\end{array}
			\right )
			\left (	\begin{array}{c}
			1 \\
			1 \\
			1 \\
			1\\
				\end{array}
			\right )
			+
			\left (	\begin{array}{c}
				0 	\\
				0	\\
				1	\\
				0	\\
				\end{array}
		\right)fsin(\omega x_4), \qquad \text{i=1,2,3},
\label{eqn:mmlc_vect_field}
\end{equation}
where we assume $a_1 = a_3$.\\
If the boundary equilibrium points at the two switching manifolds are taken as $X^\ast=\{\pm 1,x_{20},x_{30},x_{40}\}$, where $x_{20} \neq 0$, then we find that $F_2(x,\mu) \neq F_1(x,\mu)$ at $x \in \Sigma_{1,2}$ and $F_2(x,\mu) \neq F_3(x,\mu)$ at $x \in \Sigma_{2,3}$. Under these conditions the system is said to have degree of smoothness of order \emph{one}, that is $r = 1$, where $r$ is the order of discontinuity. Hence the memristive MLC circuit can be considered to behave as a \textit{Filippov} system or a \textit{Filippov Flow} capable of exhibiting \emph{sliding} bifurcations.

\subsection{Conditions for Sliding Bifurcations in Memristive MLC Circuit}

In this section the conditions for the occurrence of sliding bifurcations in the memristive MLC circuit are derived from the general conditions for the onset of various types of sliding bifurcations given in Section 2.2. Firstly the defining condition for all sliding motions in the memristive MLC circuit is 
\begin{equation}
		x_2 = 0 \qquad \qquad  \textrm{for} \qquad |x_1| < 1
	\label{eqn:mmlc_slide_cond1}
\end{equation}
This condition is to ensure that all the trajectories impinge on the switching manifold $\hat{\Sigma}$ transversely. In addition to this there is a non-degeneracy assumption valid for all sliding bifurcations 
\begin{equation}
	x_2 > 0 \qquad \qquad \textrm{for} \qquad |x_1| > 1
	\label{eqn:mmlc_slide_cond2}
\end{equation}
For \emph{crossing-sliding} and \emph{grazing-sliding} bifurcations we have an extra non-degeneracy condition, namely
\begin{equation}
	(-a_1x_2 +x_3) > 0 \qquad \qquad \textrm{for} \qquad |x_1| < 1
	\label{eqn:mmlc_slide_cond3a}
\end{equation}
for the \emph{crossing-sliding} case while
\begin{equation}
	(-a_2x_2 +x_3) > 0 \qquad \qquad \textrm{for} \qquad |x_1| < 1
	\label{eqn:mmlc_slide_cond3b}
\end{equation}
for the \emph{grazing-sliding} case.
This extra non-degeneracy condition ascertains whether the boundary $\partial \widehat{\Sigma}$ is attracting or repelling to the sliding flow.
For the \emph{switching-sliding} bifurcation, this condition becomes
\begin{equation}
	(-a_1x_2 +x_3) < 0 \qquad \qquad \textrm{for} \qquad |x_1| < 1
	\label{eqn:mmlc_slide_cond4}
\end{equation}
while for \emph{adding-sliding} bifurcation it is 
\begin{equation}
	(-a_1x_2 +x_3) = 0 \qquad \qquad \textrm{for} \qquad |x_1| < 1
	\label{eqn:mmlc_slide_cond5}
\end{equation}
This condition is a defining condition for the existence of a point of tangency of the adding-sliding flow with $\partial \widehat{\Sigma}^-$ at the bifurcation point. Further for the \emph{adding-sliding} bifurcation an extra inequality condition has to be satisfied, namely
\begin{equation}
	(1-a_1)<0.
	\label{eqn:mmlc_slide_cond6}
\end{equation}
As the parameter $a_1$ is chosen to be negative for this system, the condition given by Eq. (\ref{eqn:mmlc_slide_cond6}), cannot be satisfied. Consequently the \textit{adding-sliding} bifurcations cannot be  observed in the memristive MLC circuit.
 
\subsection{ZDM Corrections for Sliding Bifurcations in Memristive MLC Circuit}
The Zero Time Discontinuity Map (ZDM) corrections for all the four types of sliding bifurcations for the memristive MLC circuit described by Eq. (\ref{eqn:smooth_odes}) evaluated at the boundary equilibrium points $X^\ast=\{\pm 1,x_{20},x_{30},x_{40}\}$, where $x_{20} \neq 0$ are derived. For an elaborate derivation of ZDM corrections for a \textit{Filippov system}, refer \cite{dib08}. 
\begin{enumerate}
\item
{\bfseries{Crossing-sliding:}}\\
For trajectories starting in the region $S_2$,$(H(x)<0)$ the ZDM correction for {\emph{crossing-sliding}} bifurcation is 
\begin{equation}
x \mapsto \left\{
				\begin{array}{lcl}
					x & \text{if} & (x_2-x_2^*) \leq 0,\\ 
					x+\delta_{CS} & \text{if} & (x_2-x_2^*) > 0, \\
				\end{array}
			\right.
		\label{eqn:mmlc_slide_zdm1}
\end{equation}
where 
\begin{equation}
\delta_{CS}=\left \{
	\begin{array}{llll}
		0, & \dfrac{1}{2}\dfrac{(a_1-a_2)\epsilon^2}{a_1x_2}, &0, &0 \nonumber
	\end{array}
			\right \}
\end{equation}
and $\epsilon = (x_2-x_2^\ast)$.

\item
{\bfseries{Grazing-sliding:}}\\
For trajectories starting in the region $S_1$,$(H(x)>0)$ the ZDM correction for {\emph{grazing-sliding}} bifurcation  is 
\begin{equation}
x \mapsto \left\{
				\begin{array}{lcl}
					x & \text{if}   & (x_1-x_1^*) \geq 0, \\ 
					x-\delta_{GS} & \text{if} & (x_1-x_1^*) < 0, \\
				\end{array}
			\right.
		\label{eqn:mmlc_slide_zdm2}
\end{equation}
where 
\begin{equation}
\delta_{GS}=\left \{
	\begin{array}{llll}
		0, & (a_1-a_2)\epsilon, &0, &0 \nonumber
	\end{array}
			\right \}
\end{equation}
and $\epsilon = (x_1-x_1^\ast)$.
\item
{\bfseries{Switching-sliding:}}\\
For trajectories starting in the region $S_2$,$(H(x)>0)$ the ZDM correction for {\emph{switching-sliding}} bifurcation is 
\begin{equation}
x \mapsto \left\{
				\begin{array}{lcl}
					x & \text{if} & (x_2-x_2^*) \leq 0,\\ 
					x+\delta_{SS} & \text{if} & (x_2-x_2^*) > 0. \\
				\end{array}
			\right.
		\label{eqn:mmlc_slide_zdm3}
\end{equation}
where 
\begin{equation}
\delta_{SS}=\left \{
	\begin{array}{llll}
		0, & -\dfrac{2(a_1-a_2)^2x_2\epsilon^3}{3 a_1^2}, &0, &0 \nonumber
	\end{array}
			\right \}
\end{equation}
and $\epsilon = (x_2-x_2^\ast)$.
\item
{\bfseries{Adding-sliding:}}\\
For trajectories starting in the region $S_1$,$(H(x)<0)$ the ZDM correction for {\emph{adding-sliding}} bifurcation is 
\begin{equation}
x \mapsto \left\{
				\begin{array}{lcl}
					x & \text{if} & (x_2-x_2^*) \leq 0,\\ 
					x+\delta_{AS} & \text{if} & (x_2-x_2^*) > 0, \\
				\end{array}
			\right.
		\label{eqn:mmlc_slide_zdm4}
\end{equation}
where 
\begin{equation}
\delta_{AS}=\left \{
	\begin{array}{llll}
		0, & \dfrac{9}{2}\dfrac{(a_1-a_2)^2\epsilon^2}{(a_1^2-\beta)x_2}, &0, &0 \nonumber
	\end{array}
			\right \}
\end{equation}
and $\epsilon = (x_2-x_2^\ast)$.
\end{enumerate}

\begin{figure}[htbp]
	\centering
	\resizebox{0.65\textwidth}{!}
		{\includegraphics{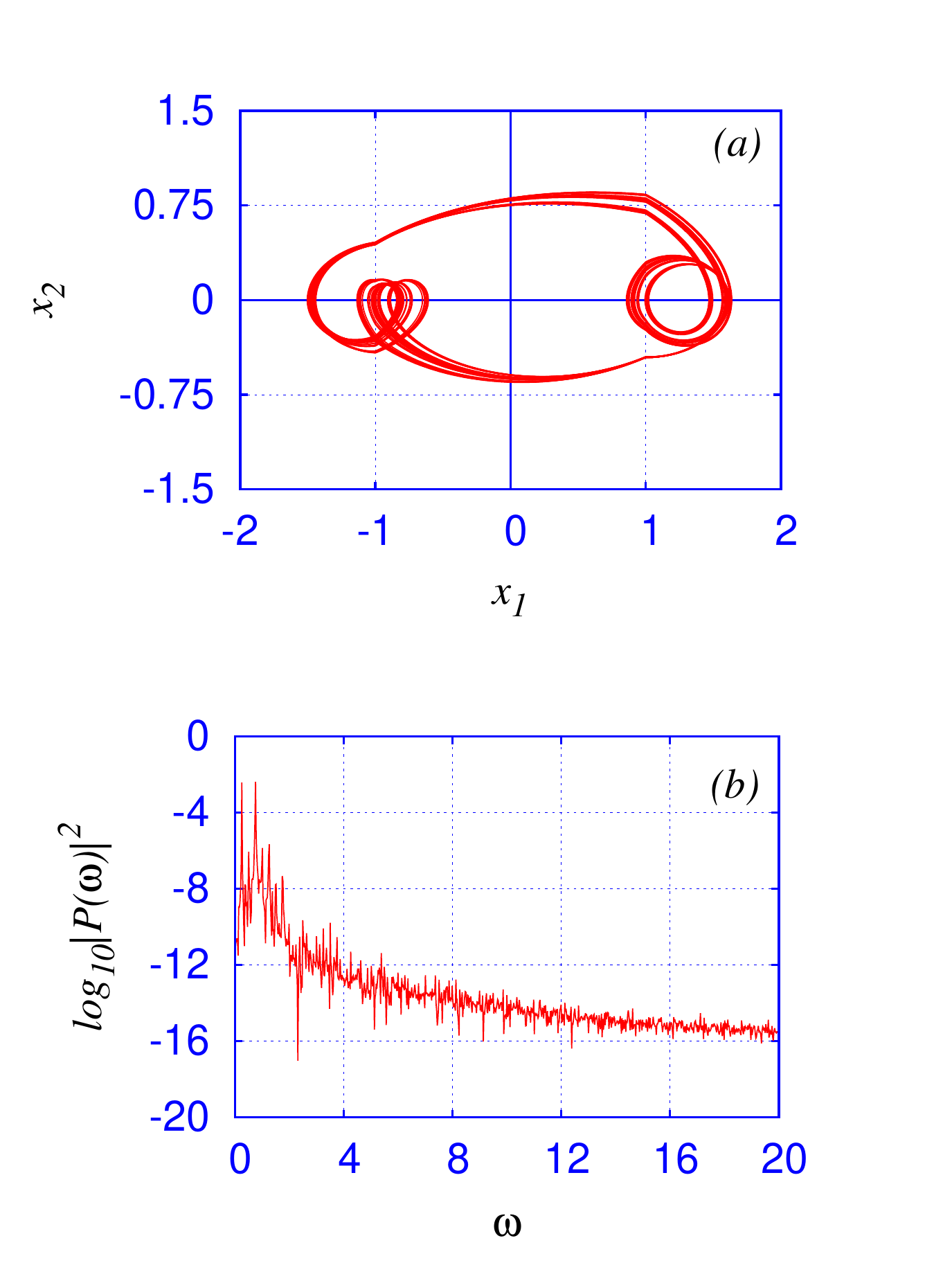}}
	\caption[Sliding Bifurcations observed in Memristive MLC Circuit] { (a) The phase portrait of the memristive MLC circuit obtained numerically for $a_{1,3} = -0.55$, $a_2 = -1.02$, $\omega = 0.75$, $f = 0.2$ and $\beta = 0.99$ showing three types of sliding bifurcations admitted by the circuit. (b) The power spectrum of the $x_2$ variable showing quasi-periodic motion.}
	\label{fig:mmlc_slide_bif1}
\end{figure}

\begin{figure}[htbp]
	\centering
	\resizebox{1.0\textwidth}{!}
		{\includegraphics{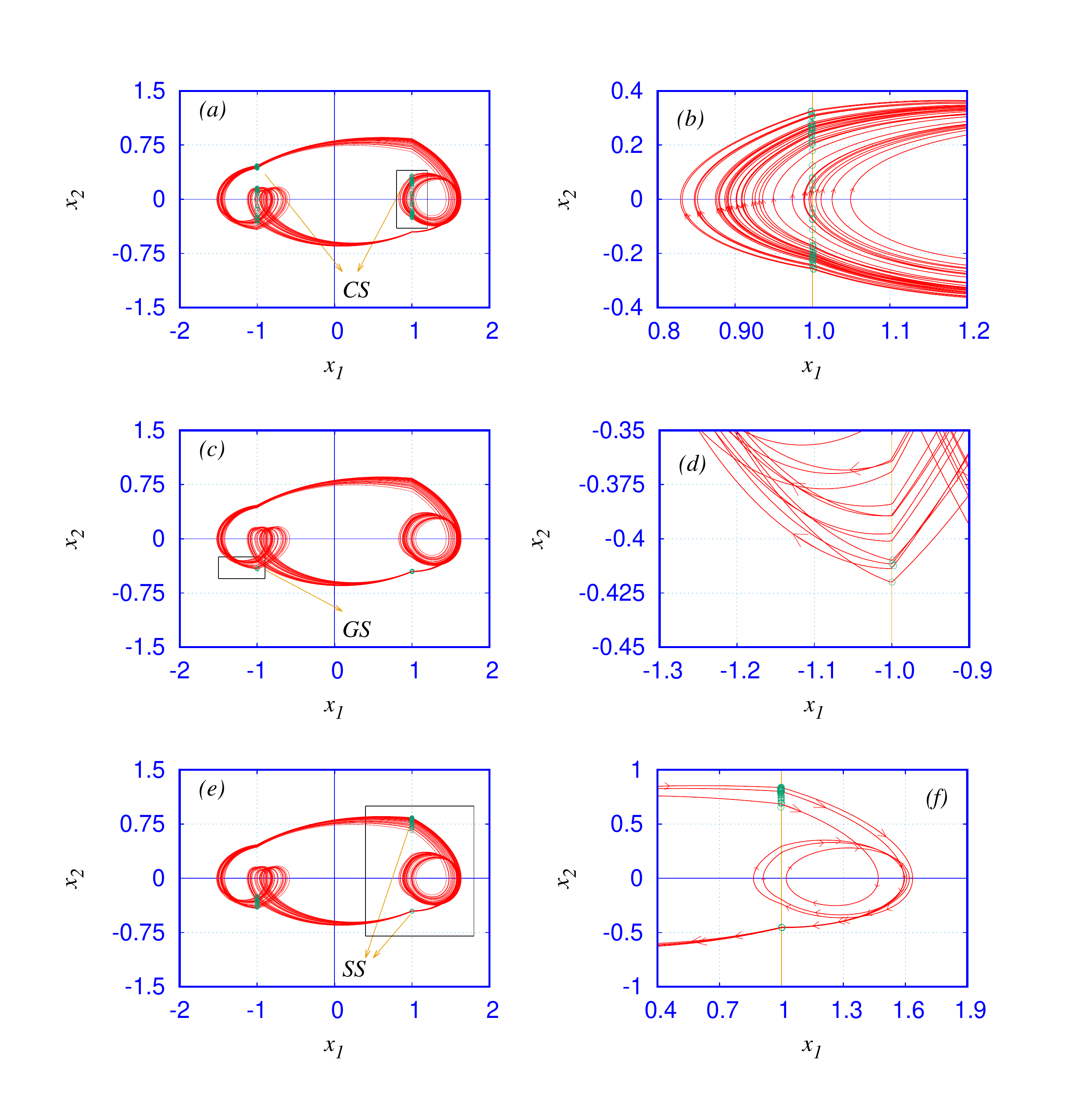}}
	\caption[Sliding Bifurcations observed in Memristive MLC Circuit] {The phase portraits in the $(x_1-x_2)$ plane capturing (a) crossing-sliding $(CS)$ bifurcations (c) grazing-sliding $GS$ bifurcations and (e) switching-sliding $SS$ bifurcations. The figures (b), (d) and (f) are the blown up portions of the same respectively, for greater clarity. The parameters are the same as mentioned in Fig. (\ref{fig:mmlc_slide_bif1}).}
	\label{fig:mmlc_slide_bif2}
\end{figure}

\begin{figure}
	\centering
	\resizebox{0.55\textwidth}{!}	
		{\includegraphics{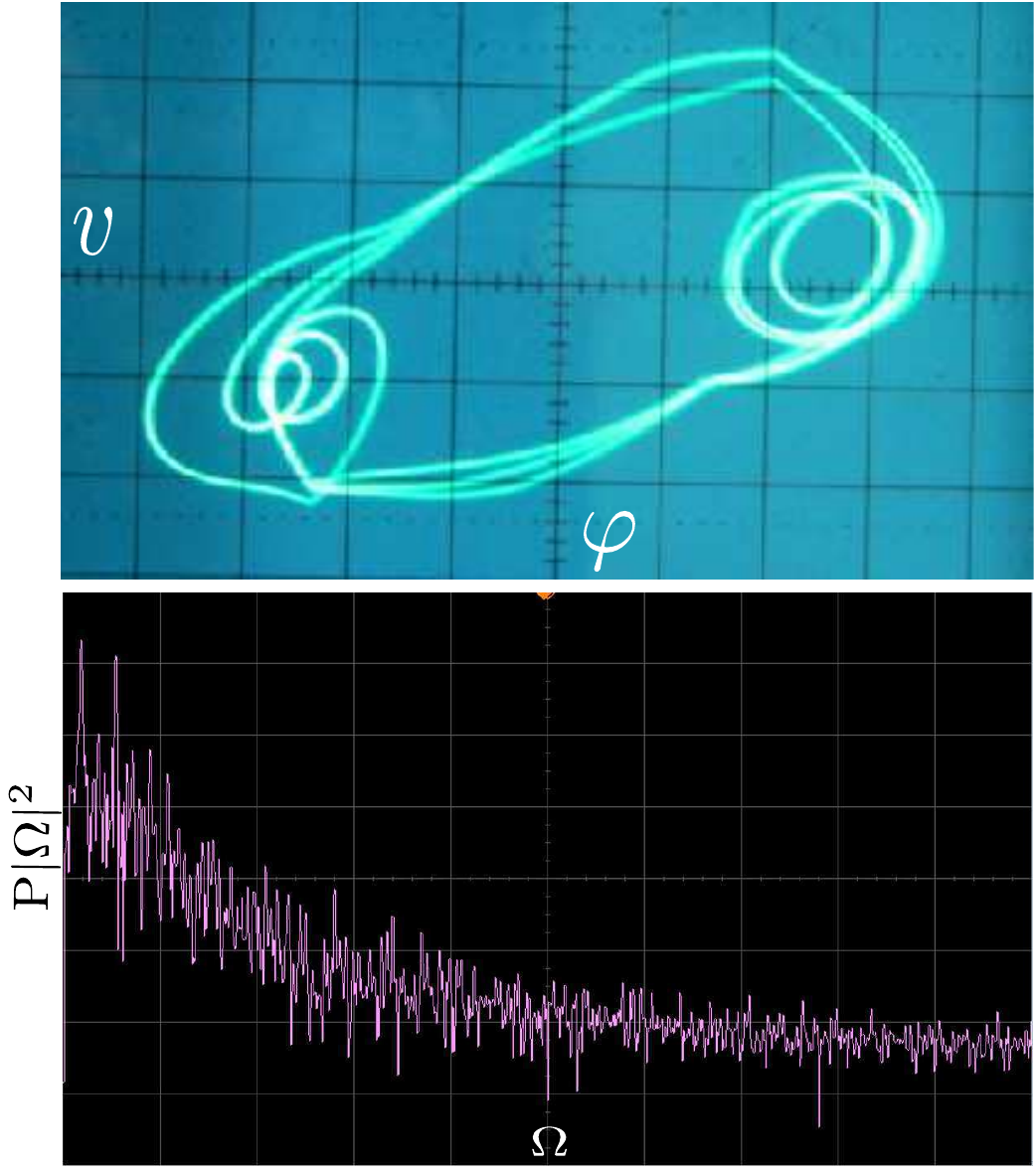}}
	\caption[Experimental Phase Portrait showing Sliding Bifurcations and its power spectrum]{ Figure (upper panel) is the experimental phase portrait in the $(\phi-v_C)$ plane showing three types of sliding bifurcations of the periodic orbits and (lower panel) the power spectrum of the voltage across the capacitor $v_c$. The values of the memristive part of the circuit are kept constant as in Fig. \ref{fig:mem_pm}, while the other parameters are fixed as  $L = 50 mH$, $C = 30 nF$, $F = 1.390 V Vpp$ and frequency $f = 1.0744 KHz$. The phase portrait corresponds well to Fig. \ref{fig:mmlc_slide_bif1} (a) while the power spectrum is in good agreement with its numerical equivalent shown in Fig. \ref{fig:mmlc_slide_bif1} (b).} 
	\label{fig:expt_mmlc_sl_spec1}
\end{figure}

\section{Sliding Bifurcations Induced Dynamics of the Memristive MLC Circuit}
The memristive MLC circuit admits three types of sliding bifurcations, namely the crossing-sliding, grazing-sliding and switching sliding bifurcations, for proper choice of circuit parameters. The dynamics has been studied by numerical simulation using RK4 algorithm. The normalized parameters are chosen as $a_1 = -1.02$, $a_2 = -0.55$, $\omega = 0.75$ and $f = 0.2$. Here $\beta$ is taken as the control parameter. As this control parameter is varied in the range $0.7 < \beta < 1.2$ we find that the circuit exhibits either one or two or all the three types of sliding bifurcations mentioned above. These are observed after the conditions for the occurrence of sliding bifurcations and the necessary Zero Discontinuity Map corrections derived above are applied to the trajectories. The sliding bifurcations so observed for $\beta = 0.99$ are shown in Fig. \ref{fig:mmlc_slide_bif1}(a). The power spectrum of the $x_2$ variable pointing to a multiple periodic behaviour is shown in Fig. \ref{fig:mmlc_slide_bif1}(b).

\begin{figure}[htbp]
	\centering
	\resizebox{0.65\textwidth}{!}
		{\includegraphics{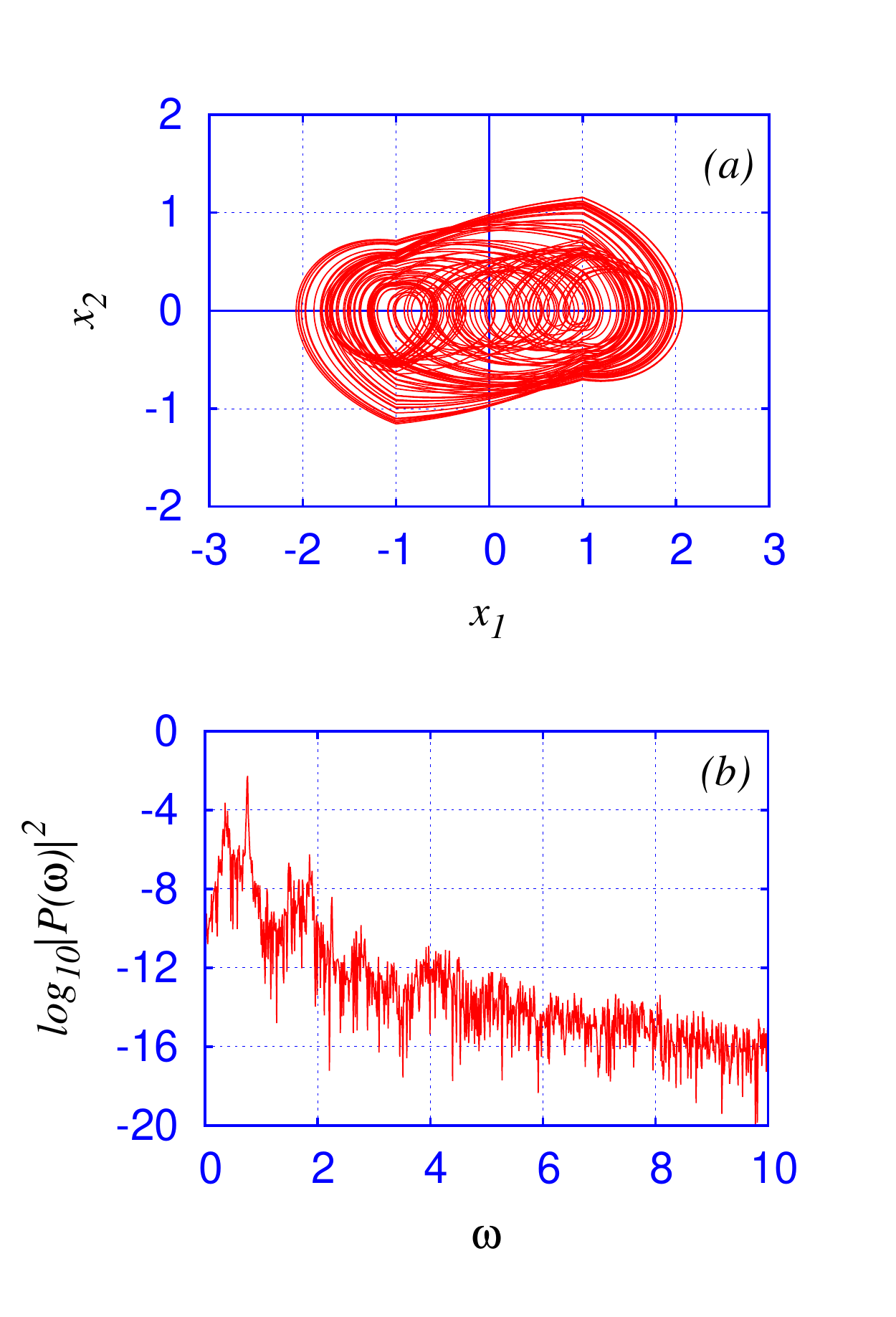}}
	\caption[Sliding Bifurcations induced Chaos in Memristive MLC Circuit] {(a)  The phase portrait of the chaotic attractor arising due to repeated sliding bifurcations occurring at the switching boundaries $\Sigma_{1,2}$ and $\Sigma_{2,3}$ for the parametric values  $a_{1,3} = -0.55$, $a_2 = -1.02$, $\omega = 0.65$, $f = 0.2$ and $\beta = 0.97$ and (b) the power spectrum corresponding to the attractor in (a).}
	\label{fig:mmlc_slide_bif3}
\end{figure}

The figures in the plots \ref{fig:mmlc_slide_bif2}(a), \ref{fig:mmlc_slide_bif2}(c) and \ref{fig:mmlc_slide_bif2}(e) are the phase portraits of the system for the same set of parameters mentioned above, but highlighting the particular type of sliding bifurcations. In Fig. \ref{fig:mmlc_slide_bif2}(a), we find that the crossing-sliding of periodic orbits is highlighted while in \ref{fig:mmlc_slide_bif2}(c) and \ref{fig:mmlc_slide_bif2}(e) the grazing-sliding and switching-sliding bifurcations respectively, are clearly shown. The insets in each of these figures are expanded for greater clarity and are shown in Figs. \ref{fig:mmlc_slide_bif2}(b), \ref{fig:mmlc_slide_bif2}(d) and \ref{fig:mmlc_slide_bif2}(f) respectively.\\

The experimental observations of the sliding bifurcations induced dynamics in the memristive MLC circuit are also made. For this purpose, the parametric values of the memristive part of the circuit are kept unchanged through out the study, while the other parameters are fixed as  $L = 50 mH$, $C = 30 nF$, $F = 1.390 V Vpp$ and frequency $f = 1.0744 KHz$. The experimental phase portrait in the $(\phi-v_C)$ plane given in Fig. \ref{fig:expt_mmlc_sl_spec1}(a) depicts the three types of sliding bifurcations of the periodic orbits. This agrees qualitatively well with the sliding bifurcations shown in the numerical phase portrait in the $(x_1-x_2)$ plane given in Fig. \ref{fig:mmlc_slide_bif1}(a). The power spectrum of the voltage across the capacitor $v_c$, shown in Fig. \ref{fig:expt_mmlc_sl_spec1}(b) is in good agreement with its numerical equivalent shown in Fig. \ref{fig:mmlc_slide_bif1}(b). 

\begin{figure}
	\centering
	\resizebox{0.575\textwidth}{!}
		{\includegraphics{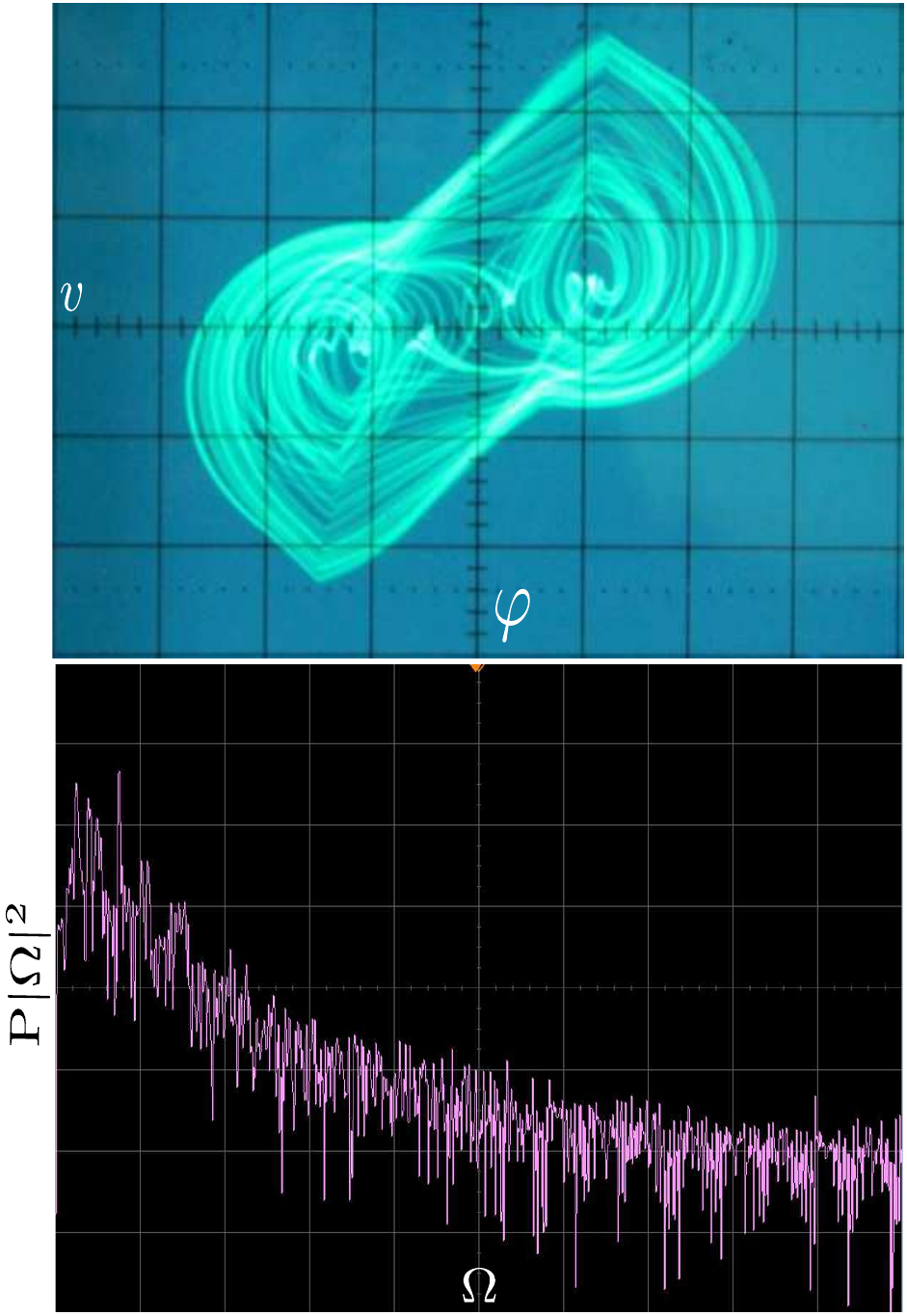}}
	\caption[Experimental Phase Portrait showing Sliding Bifurcations and its power spectrum]{Figure (upper panel) showing the experimental phase portrait in the $(\phi-v_C)$ plane exhibiting three types of sliding bifurcations of the periodic orbits and (lower panel) the power spectrum of the voltage across the capacitor $v_c$. The values of the memristive part of the circuit are kept constant as in Fig. \ref{fig:mem_pm}, while the other parameters are fixed as  $L = 50 mH$, $C = 30 nF$, $F = 1.390 V Vpp$ and frequency $f = 1.0744 KHz$. The phase portrait corresponds well to Fig. \ref{fig:mmlc_slide_bif3} (a) while the power spectrum is in good agreement with its numerical equivalent shown in Fig. \ref{fig:mmlc_slide_bif3} (b).} 
	\label{fig:expt_mmlc_sl_spec2}
\end{figure}
\subsection{Sliding Bifurcation Induced Chaos}
Repeated sliding bifurcations at the discontinuity boundaries $\Sigma_{1,2}$ and $\Sigma_{2,3}$ are found to give rise to chaotic state for the memristive MLC circuit. For example for the particular choice of parameters,  $a_{1,3} = -0.55$, $a_2 = -1.02$, $\omega = 0.65$, $f = 0.2$ and $\beta = 0.97$, the circuit is found to exhibit chaos. The phase portrait in the $(x_1-x_2)$ plane and the power spectrum of the chaotic attractor arising due to sliding bifurcations are shown in Figs. \ref{fig:mmlc_slide_bif3}(a) and \ref{fig:mmlc_slide_bif3}(b) respectively.

The experimental phase portrait of sliding bifurcations in the $(\phi-v_C)$ plane is shown in the upper panel of Fig. \ref{fig:expt_mmlc_sl_spec2}, while the power spectrum of the voltage across the capacitor $v_c$ is shown in the lower panel of Fig. \ref{fig:expt_mmlc_sl_spec2}. The values of the memristive part of the circuit are kept constant as in Fig. \ref{fig:mem_pm}, while the other parameters are fixed as  $L = 50 mH$, $C = 30 nF$, $F = 1.390 V Vpp$ and frequency $f = 1.0744 KHz$. The phase portrait corresponds well to Fig. \ref{fig:mmlc_slide_bif3} (a) while the power spectrum is in good agreement with its numerical equivalent shown in Fig. \ref{fig:mmlc_slide_bif3} (b). 

\section{Memristive Driven Chua Oscillator}
In this section, we describe the dynamics of a driven memristive Chua's circuit. This circuit itself is obtained by modifying the driven Chua's circuit, a fourth order non-autonomous circuit first introduced by Murali and Lakshmanan in the year 1990 \cite{murali90}. The reason for selecting this circuit is that it is found to exhibit a large variety of bifurcations such as period adding,  quasi-periodicity, intermittency, equal periodic bifurcations, re-emergence of double hook and double scroll attractors, hysteresis and coexistence of multiple attractors, besides the standard bifurcations. Its dynamics in the environment of a sinusoidal excitation has been extensively  studied in a series of works by Murali and Lakshmanan \citep{murali90,murali91,murali92a,murali92b,murali92c}. Due to its simplicity and the rich  content of its nonlinear dynamical phenomena, the driven Chua's circuit  continues to evoke renewed interest by researchers in the field of non-linear electronics \citep{anish93,zhu97,elwakil02,srini09} 

\begin{figure}[htbp]	
	\centering
		\resizebox{0.6\textwidth}{!}	
		{\includegraphics{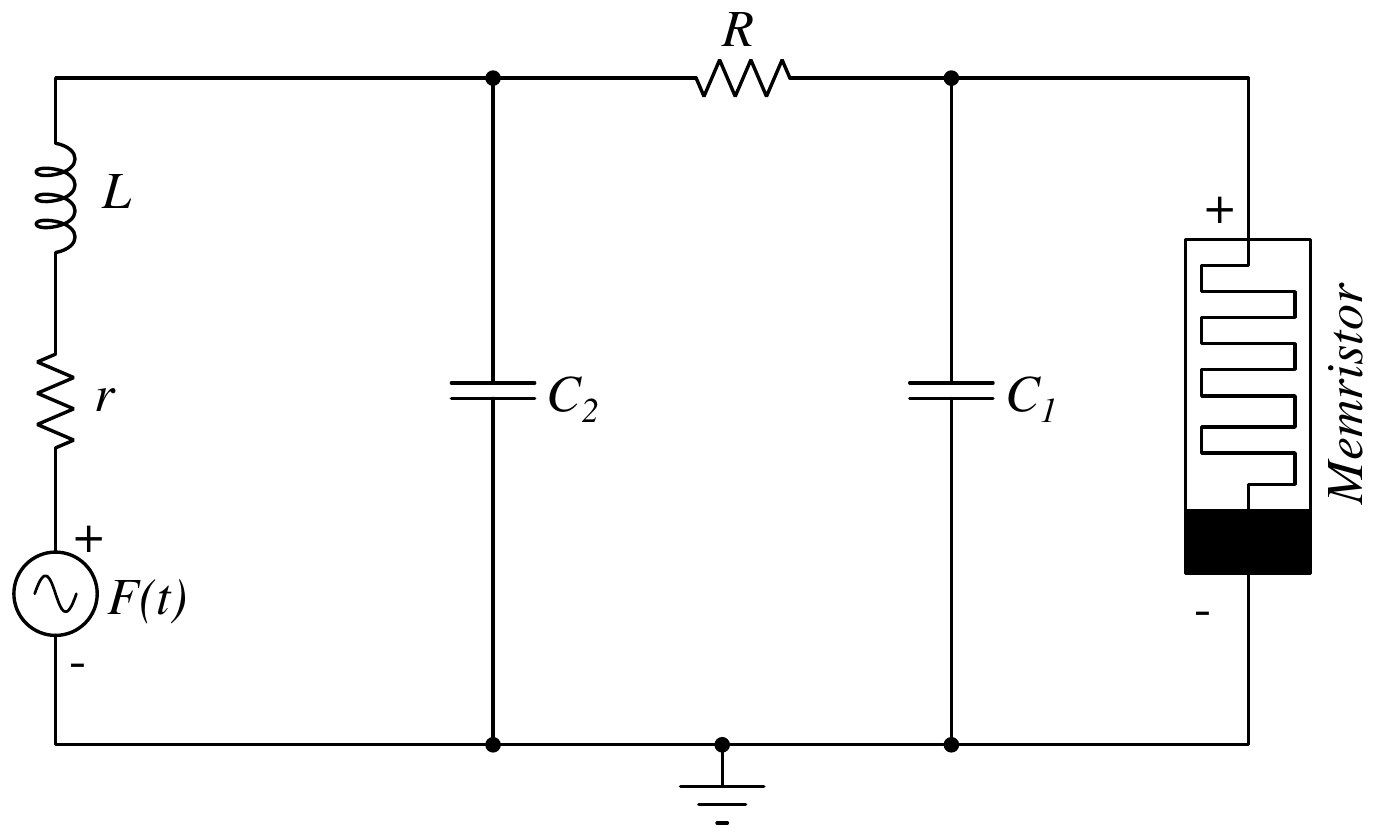}}
	\caption[Driven Memristive Chua's Circuit]{The experimental realization of the driven memristive Chua's circuit. }
	\label{fig:driv_mchua_ckt}
\end{figure}

The driven memristive Chua's circuit is obtained by replacing the Chua's diode in the original driven Chua's circuit with the active flux controlled active memristor introduced by the authors in 2011, as its nonlinearity. Further the inductances in parallel in the Driven Chua's circuit are replaced by a single equivalent inductance. The experimental realization of the driven memristive Chua's circuit is shown in Fig. \ref{fig:driv_mchua_ckt}.  
The circuit equations, obtained by applying Kirchhoff's laws are 
\begin{eqnarray}
\frac{d\phi}{dt}  & = & v_1, \nonumber \\
C_1\frac{dv_1}{dt}  & = & \frac{v_2 - v_1}{R} - W(\phi)v_1,  \nonumber \\ 
C_2\frac{dv_2}{dt}  & = & \frac{v_1 - v_2}{R} + i_L, \nonumber\\
L \frac{di_L}{dt}  & = &  -v_2 -ri_L +F\sin(\Omega t).
	\label{eqn:Mem_Chua}
\end{eqnarray}
Here $ W(\phi)$ is the memductance of the memristor, defined as in Eq. (\ref{eqn:W}) 

The circuit equations (\ref{eqn:Mem_Chua}) are normalized using the rescaling parameters
\begin{eqnarray}
x_1 = \frac{\phi}{RC_2}, x_2 = v_1, x_3 = v_2, x_4 = i_LR, \nonumber \\
\alpha = \frac{C_2}{C_1}, \beta = \frac{R^2C_2}{L}, \gamma = \frac{RC_2r}{L}, f = \beta F \equiv \frac{R^2C_2 F}{L} , \omega = \Omega R C_2.
	\label{eqn:Mem_Chua_Rescale}
\end{eqnarray}
The normalized equations so obtained are
\begin{eqnarray}
\dot{x_1}  & = & x_2, \nonumber \\
\dot{x_2}  & = & \alpha \left( x_3-x_2-W(x_1)x_2\right) , \nonumber \\ 
\dot{x_3}  & = & x_2 - x_3 + x_4,\nonumber \\ 
\dot{x_4}  & = & -\beta x_3 -\gamma x_4 -f \sin(\omega t).
	\label{eqn:Mem_chua_Nor}
\end{eqnarray}
Here dot stands for differentiation with respect to t. The normalized memductance of the memristor $W(x_1)$ is the same as that defined in Eq. (\ref{eqn:W_nor}).

\subsection{Memristive Driven Chua Oscillator as a Non-smooth System}

Similar to the memristive MLC circuit, the driven memristive Chua's oscillator can be considered as a piecewise-smooth system by virtue of the discontinuous nature of its nonlinearity, namely the memristor. The system possesses two discontinuity boundaries $\Sigma_{1,2}$ and $\Sigma_{2,3}$ which are symmetric about the origin, caused by the switching of the memristive states. These boundaries are defined by the zero sets of the smooth function $H_1(x, \mu) = (x-x^\ast)$, $x^\ast = +1$ and $H_2 = (x-x^\ast)$, $x^\ast = -1$, respectively.  Consequently the phase space gets divided into three sub-spaces $S_1$, $S_2$ and $S_3$. In the nonsmooth framework outlined earlier, the driven memristive Chua's oscillator can be recast as a set of smooth ODE's in each of these sub-spaces as,
\begin{equation}
\dot{\mathbf{x}}(t) = 
	\left\lbrace	\begin{array}{lcccl}
		F_2(\mathbf{x},\mu),& H_1(\mathbf{x}, \mu) \geq  0& \& & H_2(\mathbf{x}, \mu) \leq 0,& \mathbf{x} \in S_2,  \\
				   &					&    &                 &           \\
		F_{1,3}(\mathbf{x},\mu), &  H_1(\mathbf{x}, \mu)< 0 &\& &H_2(\mathbf{x},\mu)> 0, &\mathbf{x} \in S_{1,3},
		\end{array}
	\right.
	\label{eqn:Chua_Nonsmooth_odes}
\end{equation}
where $\mu$ denotes parameter dependence of the vector fields and the scalar functions. The vector fields $F_i$'s are
\begin{equation}
 F_i(\mathbf{x},\mu) =  
 			\left (	
 				\begin{array}{cccc}
				0	&x_2 				&0				&0		\\
				0	&-\alpha(1+a_i)x_2 	&\alpha x_3 	&0		\\
				0	&x_2 				&-x_3 			&x_4 	\\
				0	&0					&-\beta x_3 	&-\gamma x_4\\ 	
				\end{array}
			\right )
			\left (	
 				\begin{array}{c}
 				1 \\
 				1 \\
 				1 \\
 				1 \\
 				\end{array}
			\right ) 							
			+
			\left (	
				\begin{array}{c}
				0 	\\
				0	\\
				0	\\
				1	\\
				\end{array}
		\right)fsin(\omega t), \qquad \text{i=1,2,3}
	\label{eqn:Chua_Fi}
\end{equation}
with the assumption that $a_1=a_3$.
\section{Sliding Bifurcations in Memristive Driven Chua Oscillator }
In this section we report the simultaneous occurrence of two types of sliding bifurcations in the driven memristive Chua's circuit, for the same set of circuit parameters. To analyze these sliding bifurcations, the normalized state equations for the driven memristive Chua's circuit, Eqs. (\ref{eqn:Chua_Nonsmooth_odes}), are reformulated using {\emph{Utkin's Equivalent Control Method}}, refer \citet{utkin92}, as
\begin{equation}
F_S = \dfrac{F_1+F_2}{2}+\frac{F_2-F_1}{2}\beta(x).
	\label{eqn:utkins_slide} 
\end{equation}
Here it is assumed that the system flows according to a sliding vector field, $F_S$, which is the average of the two vector fields $F_1$ in region $S_1$ and $F_2$ in region $S_2$ plus an equivalent control $\beta(x)\in[-1,1]$ specified as 
\begin{equation}
\beta(x) = -\dfrac{H_xF_1+H_xF_2}{H_xF_2-H_xF_1},
	\label{eqn:utkins_beta}
\end{equation}
in the direction of the difference between the vector fields. The sliding region is given by 
\begin{equation}
\widehat{ \Sigma} := \{ x \in \Sigma : -1 \leq \beta \leq 1 \},
	\label{eqn:utkins_sigma}
\end{equation}
and its boundaries are 
\begin{equation}
\partial \widehat{\Sigma}^{\pm} := \{ x \in \Sigma: \beta = \pm 1 \}.
	\label{eqn:utkins_sigma_bound}
\end{equation}
The vector fields $F_i$'s, are the same as defined in Eq. (\ref{eqn:Chua_Fi}) 

\begin{figure}[htbp]
	\centering
	\resizebox{1.0\textwidth}{!}
		{\includegraphics{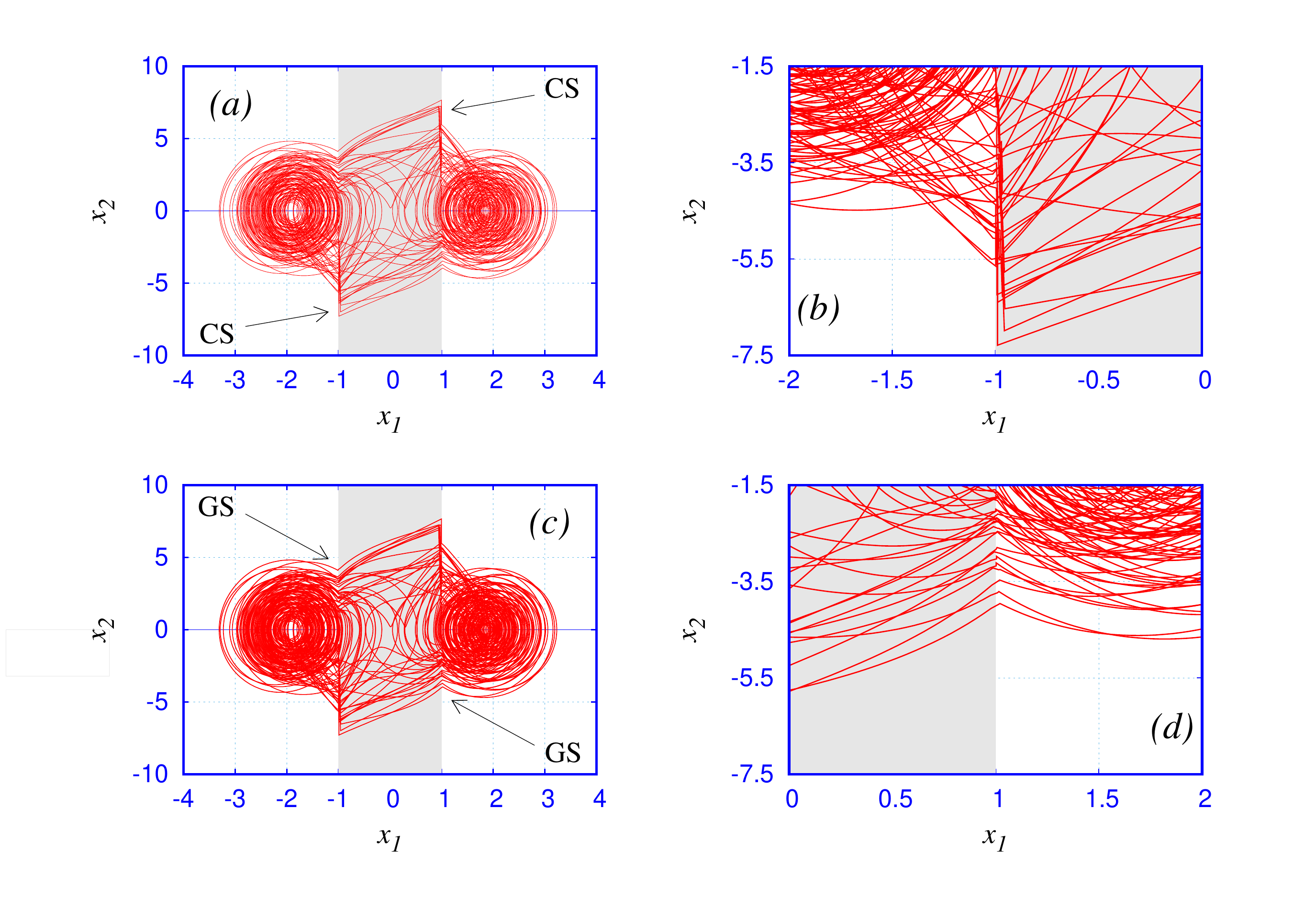}}
	\caption[Crossing-Sliding Bifurcations in Driven Memristive Chua's Circuit]{The phase portrait in the $(x_1-x_2)$ plane for $a_{1,3} = -0.68 $, $a_2 = -1.27$, $\beta=16$ and $\gamma=0.035$ ,$\omega = 0.35$, $f = 0.125$ and $\alpha=9.3$. The arrows  show (a) the {\textit{crossing-sliding}} $CS$ interactions of the trajectories  and (c) the {\textit{grazing-sliding}} $GS$ with the switching manifolds $\Sigma_{1,2}$ and  $\Sigma_{2,3}$. Figs. (b) and (d) are the enlargements of the corresponding phase portraits included here for greater clarity.}
	\label{fig:mchua_slide}
\end{figure}
\subsection{Crossing-Sliding and Grazing-Sliding Bifurcations}
Based on the conditions for the occurrence of various types of sliding bifurcations, refer Eqs. (\ref{eqn:slide_cond1})-(\ref{eqn:slide_cond6}),
the conditions for observing the same in the memristive driven Chua oscillator are derived. Using these, we find that the driven memristive Chua's circuit exhibits {\textit{crossing-sliding}} bifurcations and  {\textit{grazing-sliding}} bifurcations when the trajectories cross the switching manifolds $\Sigma_{i,j}$'s a multiple number of times as the dynamics unfolds. To our knowledge, it is for the first time that both types of sliding bifurcations are undergone by the system for the same set of parameters. These sliding bifurcations lead to chaotic behaviours.\\
The {\textit{crossing-sliding}} bifurcations are found to occur when some of the trajectories that cross the switching manifolds $\Sigma_{i,j}$'s, for $i,j=1,2$ are incident within the sliding regions $\widehat{\Sigma}_{i,j}$. In these sliding regions the trajectories show a sliding motion for a short interval of time and are governed by the sliding vector field $F_S$ before they reach the sliding boundaries $\partial \widehat{\Sigma}^{\pm}$ and move over to regions governed by either $F_2$ or $F_{1,3}$. 

The {\textit{grazing-sliding}} bifurcations are similar to the {\textit{grazing}} bifurcations encountered in general piecewise-smooth continuous flows having a higher degree of smoothness. However, these bifurcations differ from pure grazing bifurcations in that, some of the trajectories that are incident on the sliding regions $\widehat{\Sigma}_{i,j}$, exhibit sliding motion in small segments of these regions before they cross the sliding boundaries $\partial\widehat{\Sigma}^{\pm}$ and leave the switching manifolds $\Sigma_{i,j}$ transversely.\\
The conditions for the onset of \textit{crossing sliding} and \textit{grazing sliding} bifurcations in the memristive driven Chua oscillator are given below.\\
The defining condition for all sliding motions in the driven memristive Chua's circuit is
\begin{equation}
x_2^*=0, \qquad \qquad \textrm{for} \qquad |x_1|< 1.
	\label{eqn:Chua_slide_cond1}
\end{equation}
This condition is to ensure that all the trajectories impinge on the switching manifolds $\widehat{\Sigma}_{1,2}$ and $\widehat{\Sigma}_{2,3}$ transversely.
The non-degeneracy assumption valid for all sliding bifurcations in this system takes the form,   
\begin{equation}
x_2^* >0, \qquad \qquad \textrm{for} \qquad |x_1| > 1.
	\label{eqn:chua_slide_cond2}
\end{equation}
The extra non-degeneracy condition applicable to {\textit{crossing-sliding}}
bifurcations becomes 
\begin{equation}
\alpha [x_3-(1+a_2)x_2] >0, \qquad \qquad \textrm{for} \qquad -1 < x_1 < 1,
	\label{eqn:chua_slide_cond3a}
\end{equation}
while for {\textit{grazing-sliding}} bifurcations, we have
\begin{equation}
\alpha [x_3-(1+a_2)x_2] >0, \qquad \qquad \textrm{for}\qquad |x_1| > 1.
	\label{eqn:chua_slide_cond3b}
\end{equation}
These extra non-degeneracy conditions ascertain whether the sliding boundaries $\partial \widehat{\Sigma}_{\pm}$ are attracting or repelling to the sliding flow.\\
For {\textit{crossing-sliding}} bifurcations, the ZDM corrections are
\begin{equation}
x \mapsto \left\{
				\begin{array}{ll}
					x \qquad\qquad\text{if}\,\,x_2 \leq 0,\\ 
					x+\delta_1 \qquad \text{if}\,\,x_2 > 0, \\
				\end{array}
			\right.
		\label{eqn:chua_slide_zdm1a}
\end{equation}
where 
\begin{equation}
\delta_1=\dfrac{1}{2}\dfrac{(a_2-a_1)\alpha x_2}{(1+a_1)}. \nonumber
\end{equation} 
Similarly, for {\textit{grazing-sliding}} bifurcations, the ZDM corrections obtained are
\begin{equation}
x \mapsto \left\{
				\begin{array}{ll}
					x \qquad\qquad\text{if}\,\,x_{1d} \geq 0,\\ 
					x+\delta_2 \qquad \text{if}\,\,x_{1d} < 0,\\
				\end{array}
			\right.
		\label{eqn:chua_slide_zdm1b}
\end{equation}
where 
\begin{equation}
\delta_2 = (a_2-a_1)\alpha x_{1d}, \nonumber
\end{equation} 
and 
\begin{equation}
x_{1d} = x_1-x_1^*.
\end{equation}
The circuit parameters are fixed as $a_{1,3} = -0.68$, $a_2 = -1.27$, $\beta = 16$, $\gamma = 0.035$, $\omega = 0.35$, $f = 0.125$ with $\alpha$ as the control parameter. The phase portraits for these sliding bifurcations were obtained by incorporating proper ZDM corrections to the trajectories undergoing sliding bifurcations. \\
As the control parameter is varied in the range $ 8.0 < \alpha < 11.0$ we find {\textit{crossing-sliding}} bifurcations and {\textit{grazing-sliding}}  bifurcations occurring in the system. These are shown in Figs. \ref{fig:mchua_slide}(a) and \ref{fig:mchua_slide}(c) respectively, for the parameter value $\alpha = 9.3$.\\
In Fig. \ref{fig:mchua_slide}(a), we find that the trajectories undergo \textit{crossing sliding} interactions with the switching manifolds $\Sigma_{1,2}$ and $\Sigma_{2,3}$ at the points of crossings, shown by the arrows. The portrait in Figs. \ref{fig:mchua_slide}(b) are the enlargements of a portion of the phase portrait showing \textit{crossing sliding} bifurcations with the switching manifold $\Sigma_{1,2}$. 

\begin{figure}[htbp]
	\centering
		{\includegraphics[width=0.65\textwidth]{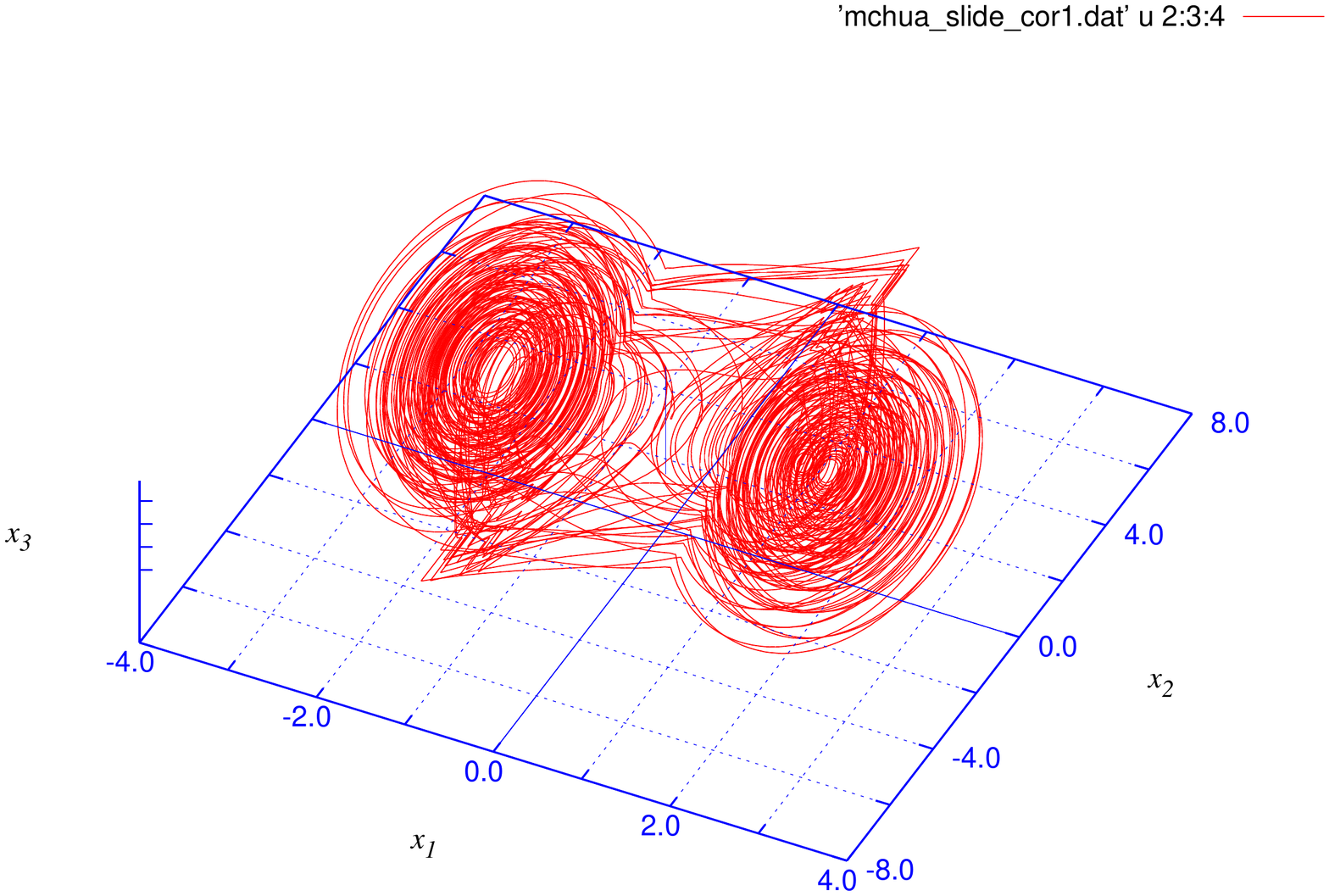}}
	\caption[Crossing-Sliding and Grazing-Sliding Bifurcations: 3D-Plot]{ 3D phase portrait of the attractor shown in Figs. \ref{fig:mchua_slide} for the same set of parameters.}
	\label{fig:mchua_slide_numeric}
\end{figure} 

\begin{figure}[htbp]
	\centering
	\resizebox{1.0\textwidth}{!}
		{\includegraphics{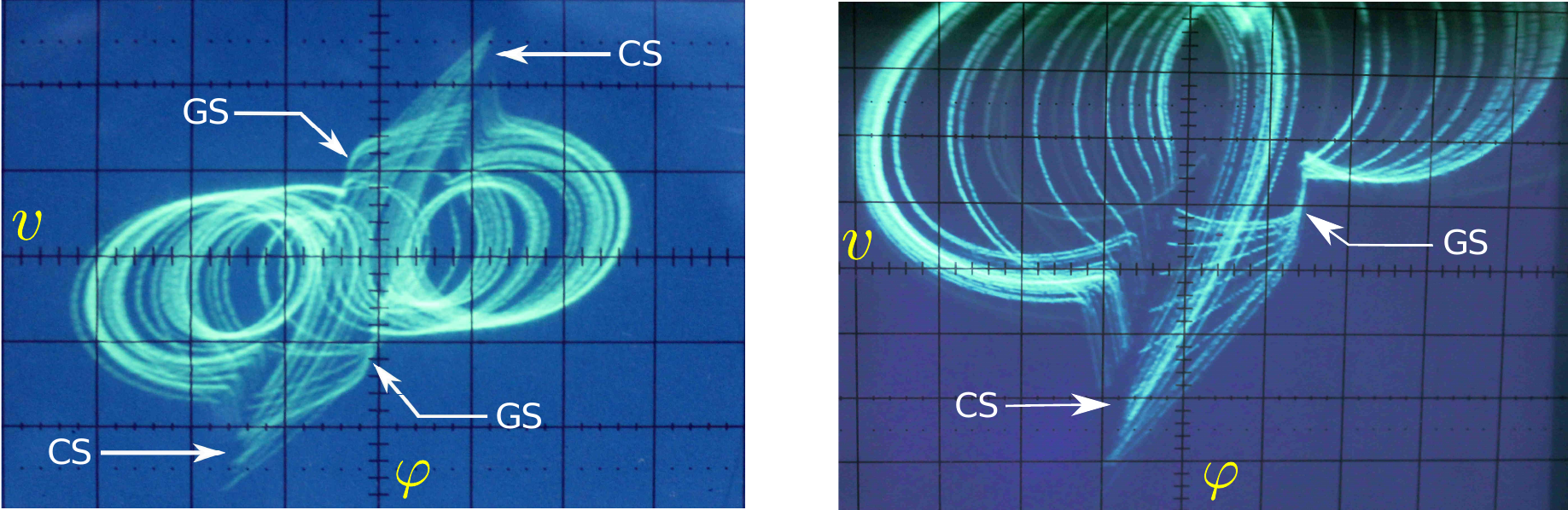}}
	\caption[Crossing-Sliding and Grazing-Sliding Bifurcations: Experimental Observations]{The phase portrait in the $(\phi - v_1)$ plane for $L = 26.76 mH$, $C_1 = 40.77 pF$, $C_2 = 98.79 nF$, $R = 740 \Omega$, strength of the external forcing as $F_o = 1.864 KHz$ and frequency as $\Omega = 1864 Hz$. The arrows  show clearly the {\textit{crossing-sliding}} $CS$ and the {\textit{grazing-sliding}} $GS$ interactions of the trajectories with the switching manifolds $\Sigma_{1,2}$ and  $\Sigma_{2,3}$.}
	\label{fig:mchua_slide_exp}
\end{figure} 
Similarly in Fig. \ref{fig:mchua_slide}(c), we find that the trajectories undergo \textit{grazing sliding} interactions with the switching manifolds $\Sigma_{1,2}$ and $\Sigma_{2,3}$ at the points of crossings, shown by the arrows. The portrait in Fig. \ref{fig:mchua_slide}(d) show the enlargements of a portion of the phase portrait showing \textit{grazing sliding} bifurcations with the switching manifold $\Sigma_{2,3}$. The 3D phase portrait of the attractor shown in Figs. \ref{fig:mchua_slide} for the same set of parameters is shown in Fig. \ref{fig:mchua_slide_numeric}.

\subsection{Experimental Verification}
The \textit{crossing-sliding} and \textit{grazing-sliding} bifurcations are verified experimentally. For this the circuit parameters are chosen as $L = 26.76$ mH, $C_1 = 40.77$ pF, $C_2 = 98.79$ nF, $R = 740 \Omega$, strength of the external forcing as $F_o = 1.864$ KHz and frequency as $\Omega = 1864$ Hz. The phase portrait obtained in the $(\phi - v_1)$ plane is shown in Fig. \ref{fig:mchua_slide_exp}(a) and a blown up portion of the same is shown in \ref{fig:mchua_slide_exp}(b) for greater clarity.

\section{Conclusion}

In this paper, we have listed the different types of sliding bifurcations exhibited by a general \textit{n-dimensional} non-smooth system and the conditions for the occurrence of the same. We have also given the analytic expressions for the zero-time discontinuity mapping corrections for realizing these numerically. Next we have described the piecewise-smooth characteristic of the active flux-controlled memristor in the $(\phi-q(\phi))$ and $(\phi-W(\phi))$ planes. Further we have obtained its $(v-i)$ characteristic using numerical simulations and hardware experiments. Though experimental studies for memristors have been reported previously, these are for memristors having smooth nonlinear characteristic. However we believe that it is for the first time that the experimental results of a memristor having nonlinear characteristic been reported.\\
 
We have then described the memristive MLC circuit and the memristive driven Chua oscillator constructed using this memristor, written their circuit equations and their normalized forms using proper rescaling parameters, converted them into \textit{Filippov} systems by proper choice of parameters, identified the sliding bifurcations exhibited by each one of them, and derived the conditions as well as discontinuity corrections for observing them numerically. Further we have studied the sliding bifurcation induced dynamics and given experimental evidence for them.\\

Our studies have shown that the three segment piecewise-linear characteristic analog memristor model that we have designed earlier, is a robust one. This model can be used successfully to convert any standard nonlinear circuit into its memristive equivalent and then can be explored for its non-smooth dynamics. This may help in realizing the electric or electronic analogue of many of the mechanical systems whose non-smooth dynamics have been well investigated. Further studies to investigate the other types of discontinuity induced bifurcations such as \textit{stick-slip} oscillations, \textit{impact} oscillations, \textit{border-collision} bifurcations and \textit{corner-collision} bifurcations etc., in these circuits can be made. We have already initiated work on these lines and will be presenting them in future works.

\nonumsection{Acknowledgments} 
\noindent
This work has been supported by DST-SERB Distinguished Fellowship Program (SB/DF/04/2017) of M.L.
\noindent
M.L acknowledges the financial support by DST-SERB Distinguished Fellowship Program (EMR/2014/001076), Government of India.
\bibliographystyle{ws-ijbc}
\bibliography{Bibliography}
\end{document}